\documentclass[twocolumn]{aastex701}

\newcommand{\Lsun}{\text{L}_\odot}
\newcommand{\Msun}{\text{M}_\odot}

\usepackage{booktabs}
\usepackage{threeparttable}
\usepackage{placeins}
\usepackage{gensymb}
\usepackage{marvosym}

\begin{document}

\title{Accretion onto the Embedded Protostar L1527 IRS: Insights from JWST NIRSpec and MIRI Observations}

\author[0009-0008-2128-6040]{W. Blake Drechsler}
\affiliation{Department of Astronomy, University of Virginia, 530 McCormick Rd., Charlottesville, VA 22904, USA}
\email{khj5gf@virginia.edu}

\author[0000-0002-6195-0152]{John J. Tobin}
\affiliation{National Radio Astronomy Observatory, 520 Edgemont Rd., Charlottesville, VA 22903, USA}
\email{jtobin@nrao.edu}

\author[0000-0002-9209-8708]{Patrick D. Sheehan}
\affiliation{National Radio Astronomy Observatory, 520 Edgemont Rd., Charlottesville, VA 22903, USA}
\email{psheehan@nrao.edu}

\author[0000-0002-4540-6587]{Leslie W. Looney}
\affiliation{Department of Astronomy, University of Illinois, 1002 West Green St., Urbana, IL 61801, USA}
\email{lwl@illinois.edu}

\author[0000-0001-7629-3573]{S. Thomas Megeath}
\affiliation{Ritter Astrophysical Research Center, Dept. of Physics and Astronomy, University of Toledo, Toledo, OH 43606, USA}
\email{tommegeath@gmail.com}

\author[0000-0001-7591-1907]{Ewine F. Van Dishoeck}
\affiliation{Leiden Observatory, Universiteit Leiden, Leiden, Zuid-Holland, NL}
\affiliation{Max-Planck Institut für Extraterrestrische Physik, Garching bei München, Germany}
\email{ewine@strw.leidenuniv.nl}

\author[0000-0002-5714-799X]{Valentin J. M. Le Gouellec}
\affiliation{Institut de Ciències de l’Espai (ICE-CSIC), Campus UAB, Can Magrans S/N, 08193 Cerdanyola del Vallès, Catalonia, Spain}
\affiliation{Institut d’Estudis Espacials de Catalunya (IEEC), c/Gran Capita, 2-4, 08034 Barcelona, Catalonia, Spain}
\email{vlegouellec@ice.csic.es}

\author[0000-0002-8963-8056]{Thomas P. Greene}
\affiliation{IPAC, California Institute of Technology, 1200 East California Boulevard, Pasadena, CA 91125, USA}
\email{tgreene@ipac.caltech.edu}

\author[0000-0001-8822-6327]{Logan Francis}
\affiliation{Leiden Observatory, Universiteit Leiden, Leiden, Zuid-Holland, NL}
\email{francis@strw.leidenuniv.nl}

\author[0000-0001-9217-3168]{R. Devaraj}
\affiliation{Dublin Institute for Advanced Studies, DIAS Headquarters, 10 Burlington Road, D04C932 Dublin, Ireland}
\email{dev@cp.dias.ie}

\author[0000-0002-6312-8525]{Martijn van Gelder}
\affiliation{Leiden Observatory, Universiteit Leiden, Leiden, Zuid-Holland, NL}
\email{vgelder@strw.leidenuniv.nl}

\author[0000-0003-1430-8519]{Lee Hartmann}
\affiliation{University of Michigan, Ann Arbor, MI 48109, USA}
\email{lhartm@umich.edu}

\author[0000-0002-9470-2358]{Lukasz Tychoniec}
\affiliation{Leiden Observatory, Universiteit Leiden, Leiden, Zuid-Holland, NL}
\email{lukasz.tychoniec@gmail.com}

\author[0000-0002-3950-5386]{Nuria Calvet}
\affiliation{University of Michigan, Ann Arbor, MI 48109, USA}
\email{ncalvet@umich.edu}

\author[0000-0002-3747-2496]{William J. Fischer}
\affiliation{Space Telescope Science Institute, 3700 San Martin Drive, Baltimore, MD 21218, USA}
\email{wjfischer@gmail.com}

\begin{abstract}

Accretion is the primary driver of protostellar evolution, regulating mass assembly and shaping the physical and chemical environments of young stellar objects. Quantifying accretion in the Class 0 protostellar phase is particularly important, yet remains observationally challenging due to high extinction toward the central protostars. In this paper, we present JWST NIRSpec and MIRI/MRS IFU data towards the Class 0 protostar L1527 IRS. We extract one-dimensional spectra and find emission from atomic and molecular hydrogen, water, OH, and several ionic species. The atomic hydrogen lines, Br$\alpha$, Pf$\alpha$, and Pf$\gamma$ are the most critical to this study since they can be used as accretion diagnostics. The existence of these atomic hydrogen lines viewed in scattered light indicates that accretion is likely occurring magnetospherically rather than through a boundary layer. Moment 0 emission maps show that the hydrogen emission is co-spatial with the scattered light continuum with a strong east-west asymmetry which is not due to outflow shocks. We additionally present moment 0 maps of other detected species and discuss their emission morphology. By primarily analyzing the Br$\alpha$ line, the strongest of our detected atomic hydrogen lines, we characterize the accretion onto L1527 IRS by estimating the accretion luminosity to be $\sim0.4~\Lsun$ and the accretion rate to be $\sim1\times10^{-7}~ \Msun \text{yr}^{-1}$. We lastly discuss the implications of our results with respect to both non-steady and asymmetric accretion possibly occurring in L1527 IRS.

\end{abstract}



\section{Introduction}
% L1527 IRS (IRAS 04368+2557; hereafter L1527)

The process of star formation begins with the gravitational collapse of a cold, dense molecular cloud core. An individual core may further fragment, giving rise to both individual protostars and multiple systems \citep{offner2023}. As a core further collapses into a protostar, a protostellar disk forms to conserve angular momentum \citep{cassen&moosman1981, terebey1984}. 

Protostars are typically categorized into different observational classes that are thought to correlate with different evolutionary stages \citep{tobin&sheehan2024}. The youngest of these are Class 0 protostars \citep{andre1993, andre2000}.  These youngest protostars are defined as still being deeply embedded in a dense envelope of material from the parent cloud/ core that has not yet been accreted onto the star. In Class 0's, the mass of this envelope exceeds that of the protostar, unlike more evolved Class I sources which have lower envelope masses since more of the material has been processed/ dispersed. During the most embedded stage, protostars accumulate a large fraction of their final mass \citep{dunham2010}, and so studying accretion in Class 0 protostars is critical for understanding how protostars gain their mass and grow. However, observations of Class 0 protostars are challenging due to the high extinctions resulting from their embedded nature. Thus, there is not currently a good understanding of how Class 0 protostars grow and the underlying mechanism driving their accretion. The James Webb Space Telescope (JWST) gives us the capability to peer through these circumstellar envelopes and study accretion in these embedded sources using unique diagnostic tracers that are not available at millimeter wavelengths. Observing these systems at infrared (IR) wavelengths allows us to diminish the impact of extinction that is present at optical wavelengths. In addition, JWST provides the required sensitivity to detect accretion tracing species. 

Observations of more evolved YSOs have established two mechanisms for disk mediated accretion: magnetospheric accretion and direct, boundary layer accretion. In magnetospheric accretion, the accreting material flows from the inner disk onto the surface of the protostar along the protostellar magnetic field \citep{hartmann1994}. The material reaches the protostar surface traveling at near free-fall velocity, thus the collision of the accreting material with the protostellar surface results in an accretion shock. This accretion mechanism is typically observed in more evolved Class I protostars and T Tauri stars \citep{bouvier2007, fiorellino2023}. However, it is possible for accretion to be powerful enough to overwhelm and crush the magnetosphere, leading to the second accretion mechanism; boundary layer accretion \citep{rodriguez&hillenbrand2022, cleaver2023}. In this scenario, the compression of the magnetosphere allows the accretion disk to extend inward all the way to the surface of the protostar. This results in the material being accreted more smoothly and thus there is no (or a much weaker) accretion shock. Non-ideal MHD simulations have shown that protostars must have magnetic fields of strength $>$ 1kG in order for the disk to be truncated and accretion to occur via the magnetosphere, otherwise, accretion occurs via a boundary layer \citep{gaches2024}.

The accretion shock that is present in the magnetospheric accretion scenario can heat up and excite the atomic hydrogen present around the protostar and the inner disk, as well as within the accretion flows from the disk to protostar. As a result, atomic hydrogen lines can be used as an accretion tracer in the case of magnetospheric accretion. Additionally, a lack of these features provides evidence for boundary layer accretion.

In Class I protostars the Br$\gamma$ and Pa$\beta$ accretion-tracing hydrogen lines are typically used \citep[eg.][]{connelley2014, fiorellino2023}. In Class I sources, the envelope has largely dissipated, leading to lower extinction and allowing these near-IR lines to be observed. However, these lines are often too extincted to be used to measure accretion in more embedded Class 0 protostars. JWST gives the opportunity for high sensitivity observations of the mid-IR hydrogen lines, Br$\alpha$, Pf$\alpha$, and Pf$\gamma$, which should be less impacted by extinction. JWST observations of other young protostars, however, show shocks and jets being traced by Br$\alpha$ \citep{Federman2024, leGouellec2025, delabrosse2024, tychoniec2024}. The presence of extended atomic hydrogen emission makes it more challenging to study accretion as it is difficult to disentangle the scattered light emission being produced by accretion, and locally produced shock-heated emission from jets.

In this study, we consider the Class 0 protostar L1527 IRS (IRAS 04368+2557; hereafter L1527). This source is located in the L1527 cloud in the Taurus star formation region at a distance of 139-141 pc determined by Gaia Data Release 2 \citep{luhman2018}. L1527 is a very well-studied Class 0 protostar and has a well-characterized detected Keplerian disk \citep{tobin2012, ohashi2014}. This disk has been shown to be well-resolved in scattered light L$^{\prime}$ band observations \citep{tobin2010}. L1527 was observed with ALMA as part of the Early Planet Formation in Embedded Disks (eDisk) survey, where its observed SED implied it was a Class 0 source with $L_{bol} = 1.3~\Lsun$ and $T_{bol} = 41$~K \citep{ohashi2023}. By performing radiative transfer (RT) modeling on Spitzer Space telescope data, \citet{tobin2008} estimated the total luminosity to be $L_{tot} = 2.75~\Lsun$. RT modeling found a disk mass of $M_{disk} = 0.0075~\Msun$ and a disk radius of $R_{disk} = 125$ au \citep{tobin2013}. Using the PV diagram of the eDisk data, \citet{van'thoff2023} estimated a protostellar mass for L1527 of $M_* \sim 0.5~\Msun$. L1527 is observed almost perfectly edge-on. This viewing geometry makes it easier to model the system's physical parameters due to eliminating the ambiguity that can be caused by intermediate inclinations, making it an ideal candidate for studying accretion. However, the direct line of site to the protostar is blocked by extinction from the edge-on disk, and thus the accretion lines must be observed in scattered light. 

We present JWST observations toward the Class 0 protostar L1527 IRS. In Section \ref{sec:obs}, we will discuss the observations and the data reduction process. In Section \ref{sec:results}, we will present the initial results, including extracted one-dimensional spectrum and emission maps of various detected species. In Section \ref{sec: analysis}, we will estimate the accretion luminosity of L1527. In Section \ref{sec:discussion}, we interpret our results and discuss the inferred accretion rate as well as implications related to non-steady and asymmetric accretion. Lastly, we conclude by summarizing our findings in Section \ref{sec:conclusion}.

\section{Observations and Data Reduction} \label{sec:obs}

The Class 0 protostar L1527 was observed with JWST using the Near Infrared Spectrograph (NIRSpec) \citep{jakobsen2022, boker2022} and Mid-Infrared Instrument (MIRI) Medium Resolution Spectrograph (MRS) \citep{rieke2015, wright2023} Integral Field Units (IFU). The observations were initially attempted on 2023 March 3; however the guide star target acquisition failed during the NIRSpec observations, and only the MIRI/MRS data were successfully obtained. Both the NIRSpec and MIRI data were successfully retaken on 2023 September 17.

For the NIRSpec observations, a medium resolution (G395M) grating was used so that the Br$\alpha$ line was visible across the entire IFU field of view, since Br$\alpha$ falls in the detector gap of the high resolution grating (G395H). We did not use any of the shorter wavelength gratings because emission toward the central protostar and disk at $\lambda < 3\mu$m is completely obscured. The NIRSpec data were taken in a 4-point dither pattern centered on the previously observed scattered light disk around L1527. Two integrations per dither, and 35 groups per integration were used for these observations, with a total time on source of 70 minutes. Leakage calibrations were not taken because the micro shutter array was not expected to fall on bright regions of emission.

Similarly, the MIRI/MRS observations were taken in a 4-point standard dither pattern centered on the scattered light disk. The original observations taken on 2023 March 3 used 5 integrations per dither, and 16 groups per integration, corresponding to 134 minutes on source. The re-done observations used 4 integrations per dither, and 16 groups per integration, corresponding to 107 minutes on source.  MIRI/MRS observations were needed to detect the Pf$\alpha$ hydrogen line at 7.46 m$\mu$ to complement the Br$\alpha$ and Pf$\gamma$ lines observed with NIRSpec. We only took MIRI/MRS observations using the long grating sub-band setting to maximize sensitivity since Pf$\alpha$ falls within the long sub-band of Channel 1 and we did not require continuous spectral coverage. The long module acquires data in all 4 channels simultaneously and their wavelength ranges and resolutions, along with NIRSpec's, are shown in Table \ref{tab:chan wavs}. Dedicated background observations were taken with MIRI/MRS on blank sky near L1527 also using a 4-point standard dither pattern. The original background observations used the same number of groups as the science exposures, but were taken with only a single integration per dither. The re-done background observations used the same parameters as the on-source observations.

\begin{table*}
\centering
\caption{Summary of the NIRSpec and MIRI observations.}
\begin{tabular}{lcccccc}
\hline\hline
\label{tab:chan wavs}
Channel & $\lambda_{min}$ & $\lambda_{max}$  & Resolving power & Pixel scale & Diffraction limit & RMS\\
& ($\mu$m) & ($\mu$m) & & (arcsec) & (arcsec) & (Jy)\\
\hline
NIRSpec IFU (G395M/F290LP) & 2.87 & 5.27 & ~1000 & 0.1 & 0.158 & $1.40\times10^{-5}$\\
MIRI/MRS ch1-long & 6.53 & 7.65 & 3100-3610 & 0.196 & 0.274 & $5.29\times10^{-5}$\\
MIRI/MRS ch2-long & 10.01 & 11.70 & 2860-3300 & 0.196 & 0.420 & $2.18\times10^{-5}$\\
MIRI/MRS ch3-long & 15.41 & 17.98 & 1980-2790 & 0.245 & 0.646 & $8.43\times10^{-5}$\\
MIRI/MRS ch4-long & 24.40 & 27.90 & 1630-1330 & 0.273 & 1.028 & $1.91\times10^{-3}$\\ 
\hline
\end{tabular}

\end{table*}

\subsection{Data Reduction}

We reduced the data using version 1.16.1 of the JWST pipeline and CRDS context files 'jwst\_1303.pmap'. When reducing the MIRI/MRS data, we manually ran the JWST pipeline on the stage 1 'uncal' data products. 
Stage 1 of the pipeline is responsible for the detector level corrections, such as correcting for the dark current and readout noise. Stage 1 results in images in units of count rate per second. Stage 2 of the pipeline is responsible for accounting for the flat field and WCS information. When running Stage 2, we edited the default settings to ensure we were additionally performing a dedicated background subtraction and a residual fringing correction. This step results in images with units of MJy sr$^{-1}$. Lastly Stage 3 of the pipeline is responsible for creating the spectral cubes and extracting 1D spectra. This resulted in 4 different MIRI spectral cubes, corresponding to the ch1-long, ch2-long, ch3-long, and ch4-long sub-bands.

In addition, we combined the data from both of our MIRI/MRS data sets in order to maximize our S/N. There was a small pointing offset between the data sets, so we first aligned both IFU cubes. To calculate this offset, we first median collapsed both IFU cubes along the spectral axis to obtain a representative image. We then performed a cross-correlation between these images to calculate their pixel offset. We averaged the two data sets together, using the offset to ensure they were properly aligned. In regions where the two data sets did not spatially overlap, we used the data from individual observations in order to retain the full field of view. Although we used the same offset for each spectral channel, we performed this alignment and combination separately for each of our MIRI channels.

To reduce the NIRspec data, we used the same JWST pipeline version and CRDS context file as we used for the MIRI/MRS reduction. For this data, we found we did not need to change the default settings for running any of the pipeline stages.

For MIRI MRS, the absolute flux uncertainty is estimated to be less than $0.5\%$ for channels 1-3 and within $5\%$ for channel 4 long \footnote{https://jwst-docs.stsci.edu/jwst-calibration-status/miri-calibration-status/miri-mrs-calibration-status}. The NIRSpec absolute flux uncertainty is also less than $5\%$\footnote{https://jwst-docs.stsci.edu/jwst-calibration-status/nirspec-calibration-status/nirspec-ifu-calibration-status}. We do not factor these in our reported uncertainties, instead we only report the statistical uncertainties.

\section{Results} \label{sec:results}
\subsection{Continuum Morphology}
In Figure \ref{fig:full_spec}, we present the median-collapsed spectral cubes corresponding to each channel in order to show the morphology of the continuum emission. In the continuum maps for NIRSpec and MIRI ch1 and ch2-long, we see the scattered light from the eastern and western disk surfaces. These two bright regions of emission are separated by a dark lane due to extinction from the dense disk mid-plane. We observe this optically thick dark lane due to the edge-on viewing geometry of the source. 

For the longer wavelength channels, MIRI ch3 and ch4-long, the separation between the two disk surfaces is no longer clear due to the lower spatial resolution of the data. Additionally, at these longer wavelengths, the continuum begins to become more dominated by thermal emission from the disk and inner envelope rather than scattered light from the protostar.

\subsection{Spectral Extraction}\label{sec: spec extract}
To extract one-dimensional spectra from our NIRSpec, and MIRI data cubes, we used two elliptical apertures (one for the eastern disk surface and one for the western disk surface) rather than encompassing the entire source with a circular aperture. This increases our S/N by omitting pixels with no signal due to the dark lane caused by the disk mid-plane. Additionally, this allows us to examine the spectra from the eastern and western disk surfaces separately, enabling us to search for any spatial variations. However, for MIRI ch3 and ch4-long, where the dark lane is not resolved, we instead use a circular aperture to ensure we are including all of the flux. 

\begin{figure*}
	\centering
	\includegraphics[width=\textwidth, trim={2cm, 1cm, 2cm, 1cm}, clip]{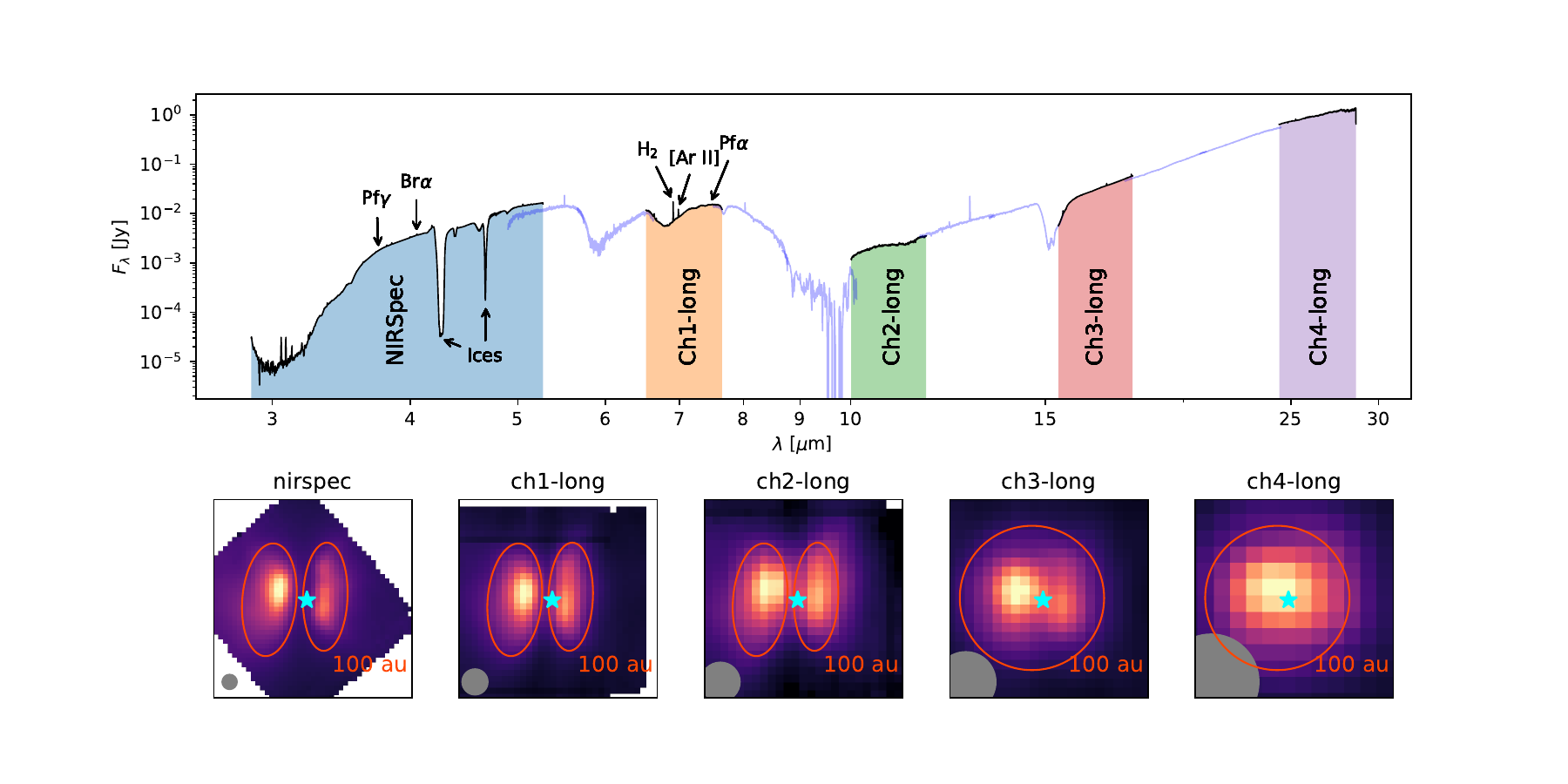}
	\caption{Our extracted 1D spectrum of L1527 using NIRSpec and MIRI data. The various observing channels are highlighted with different colors. Below the spectrum we show the median collapsed spectral cubes to demonstrate the morphology of the continuum emission for each channel. We also plot the apertures we used to extract the 1D spectra from each channel as orange ellipses. The ALMA continuum position of the source is shown with the cyan star and the diffraction limited beam is shown as a grey circle (in future Figures as well). We fill out our spectral coverage by showing data from the JOYS program (in blue) presented in Devaraj et al. (Submitted)}
	\label{fig:full_spec}
\end{figure*}

We determined our apertures by eye using the MIRI/MRS channel 1-long data. We converted these aperture into sky-coordinates, using the \texttt{photutils} python package \citep{photutils}, so that our apertures would be consistent across the other MIRI channels and NIRSpec. The coordinates of the apertures are shown in Table \ref{tab: apertures}, and the apertures are shown as contours on the continuum maps in Figure \ref{fig:full_spec}. We observe that our apertures transition from being much larger than the diffraction limited beam at the shortest wavelengths to being of a comparable size at the longest wavelengths.

To extract our spectra, we simply sum the flux within the aperture(s) for each channel in the IFU data cube. Our extracted spectrum is shown in the top row of Figure \ref{fig:full_spec}. A full MIRI-MRS spectrum taken at a somewhat lower sensitivity and slightly offset position by the JOYS program was presented in \citet{vanGelder2024} and Devaraj et al. (Submitted), which we also show in Figure \ref{fig:full_spec} to fill out our missing spectral coverage.

\begin{table*}
\centering
\caption{Apertures used for spectral extraction. For the circular aperture, the width represent the radius.}
\label{tab: apertures}
\begin{tabular}{llllll}
\toprule
aperture & RA & DEC & width & height & angle \\
& (h:m:s) & (d:m:s) & arcsec & arcsec & deg \\
\midrule
eastern & 4:39:53.94 & 26:03:09.44 & 1.17 & 2.44 & -5.00 \\
western & 4:39:53.85 & 26:03:09.50 & 0.96 & 2.34 & -2.00 \\
circular & 4:39:53.90 & 26:03:09.47 & 1.56 & - & - \\
\bottomrule
\end{tabular}
\end{table*}

\subsection{1D Spectrum}
The spectrum shown in Figure \ref{fig:full_spec} displays a rising continuum towards longer wavelengths, which is the expected behavior for protostars. In addition, we observe several clear ice absorption features, such as CO$_2$, CO, and HDO \citep{slavicinska2025}, but the discussion of these solid-state features is beyond the scope of this paper.

\subsection{Spectral Features}
We detected numerous emission lines in our spectra which are summarized in Table \ref{tab: line table}. In addition to the accretion tracing atomic hydrogen lines, we also observe several pure rotational H$_2$ lines, water, OH, and atomic ions such as [Fe II], [Ar II], and [Ne III].

\begin{table*}
\centering
\begin{threeparttable}
\caption{Summary of detected species}
\label{tab: line table}
\begin{tabular}{lrlr}
\toprule
Line & Wavelength & Structure & $^\dagger$Line Flux \\
& ($\mu$m) & & (erg s$^{-1}$ cm$^{-2}$) \\
\midrule
H$_2$ 0-0 S(22) & 3.366 & Wide angle outflow & $(2.83 \pm 0.17) \times 10^{-17}$ \\
H$_2$ 0-0 S(15) & 3.626 & Wide angle outflow & $(1.14 \pm 0.05) \times 10^{-16}$ \\
H$_2$ 0-0 S(14) & 3.724 & Wide angle outflow & $(9.86 \pm 0.18) \times 10^{-17}$ \\
Pfund gamma & 3.741 & Disk Surface & $(7.43 \pm 0.84) \times 10^{-17}$ \\
H$_2$ 0-0 S(13) & 3.846 & Wide angle outflow & $(3.75 \pm 0.04) \times 10^{-16}$ \\
H$_2$ 0-0 S(12) & 3.996 & Wide angle outflow & $(2.13 \pm 0.07) \times 10^{-16}$ \\
Bracket alpha & 4.052 & Disk Surface & $(2.40 \pm 0.15) \times 10^{-16}$ \\
H$_2$ 0-0 S(8) & 5.053 & Wide angle outflow & $(1.10 \pm 0.02) \times 10^{-15}$ \\
% $^*$H$_2$ 0-0 S(6) & 6.109 & Wide angle outflow & $(4.61 \pm 1.62) \times 10^{-16}$ \\
$[$Ni II$]$ J 3/2-5/2 & 6.636 & Narrow angle outflow & $(3.58 \pm 0.13) \times 10^{-16}$ \\
H$_2$O 6.64 & 6.642 & Disk Surface & $(3.17 \pm 0.08) \times 10^{-16}$ \\
H$_2$O 6.67 & 6.672 & Disk Surface & $(2.60 \pm 0.15) \times 10^{-16}$ \\
$[$Fe II$]$ J 3/2-5/2 & 6.721 & Narrow angle outflow & $(3.00 \pm 0.20) \times 10^{-16}$ \\
H$_2$O 6.83 & 6.826 & Disk Surface & $(8.04 \pm 0.73) \times 10^{-17}$ \\
H$_2$O 6.85 & 6.857 & Disk Surface & $(1.45 \pm 0.15) \times 10^{-16}$ \\
H$_2$O 6.86 & 6.864 & Disk Surface & $(1.80 \pm 0.10) \times 10^{-16}$ \\
H$_2$ 0-0 S(5) & 6.910 & Wide angle outflow & $(4.30 \pm 0.02) \times 10^{-15}$ \\
$[$Ar II$]$ J 1/2-3/2 & 6.985 & Narrow angle outflow & $(2.25 \pm 0.02) \times 10^{-15}$ \\
Pfund alpha & 7.460 & Disk Surface & $(3.25 \pm 0.18) \times 10^{-16}$ \\
$^*$H$_2$ 0-0 S(4) & 8.025 & Wide angle outflow & $(1.97 \pm 0.23) \times 10^{-15}$ \\
$^*$H$_2$ 0-0 S(3) & 9.665 & Wide angle outflow & $(3.46 \pm 0.77) \times 10^{-16}$ \\
$[$Ni II$]$ J 7/2-9/2 & 10.682 & Narrow angle outflow & $(9.83 \pm 1.10) \times 10^{-17}$ \\
$^*$H$_2$ 0-0 S(2) & 12.279 & Wide angle outflow & $(2.75 \pm 0.43) \times 10^{-16}$ \\
$[$Ne III$]$ J 1-2 & 15.555 & Narrow angle outflow & $(7.43 \pm 0.04) \times 10^{-16}$ \\
OH 16.02 & 16.021 & Disk Surface & $(8.42 \pm 0.88) \times 10^{-17}$ \\
OH 16.05 & 16.045 & Disk Surface & $(6.79 \pm 0.46) \times 10^{-17}$ \\
OH 16.84 & 16.840 & Disk Surface & $(1.01 \pm 0.14) \times 10^{-16}$ \\
H$_2$ 0-0 S(1) & 17.035 & Wide angle outflow & $(6.40 \pm 0.08) \times 10^{-16}$ \\
$[$Fe II$]$ J 7/2-9/2 & 17.936 & Narrow angle outflow & $(1.68 \pm 0.02) \times 10^{-15}$ \\
$[$S I$]$ J 1-2 & 25.249 & Narrow angle outflow & $(4.07 \pm 0.22) \times 10^{-15}$ \\
\bottomrule
\end{tabular}
\begin{tablenotes}
\item Note: $^*$ denotes lines that were outside of our spectral coverage and were taken from the JOYS data. The JOYS spectrum where extracted using apertures that where determined by eye based on the Ch1-long data, identical to how we extracted spectrum from our data. \\
$^\dagger$ This line flux is measured over the entire FOV, which changes as a function of wavelength.
\end{tablenotes}
\end{threeparttable}
\end{table*}

To investigate these spectral features, we first created separate spectral cubes for each line. This allows us to more easily fit the local continuum emission to each line. 
We created these spectral line cubes by simply taking a spectral slab centered on the feature of interest with a spectral width of four times the linewidth, which we estimated for each line by eye. This window size allowed us to include the local continuum emission while minimizing contamination from nearby lines.

We next estimated the continuum for each of these spectral line cubes in order to create a continuum subtracted spectral cube. To model the continuum, we considered the 1D spectrum associated with each spatial pixel position. We performed an initial smoothing of this spectrum by first masking out all lines within the spectral slab then using a median filter with a kernel size of 3 channels. We then fit this smoothed, line-free spectrum with a third order polynomial. This results in a local continuum cube that we then simply subtract from the original spectral cube to obtain our continuum subtracted cube.

\subsubsection{Line Profiles and Fluxes}
We examined the line profiles of each of our detected species using the continuum-subtracted spectral data cubes. Since many of the species have extended emission, we did not use the previously determined apertures to measure the line flux since they were based on the location of the continuum emission. Instead, we masked out the areas without line emission then summed the entire masked continuum subtracted cube to obtain a line profile. We identified the line emission by first constructing a moment 0 map then smoothing it with a Gaussian kernel. We then created our line emission mask by identifying pixels with a flux above the 0.75$\sigma$ level. We fit a 1D Gaussian to the line profile for each species then compute the line flux via: 

$$S_{\nu} = A \sigma_{\nu} \sqrt{2\pi},$$
where $S_{\nu}$ is the line flux, $A$ is amplitude of the Gaussian, and $\sigma_{\nu}$ is the standard deviation of the Gaussian in frequency units. When we perform the fitting, we calculate $\sigma$ in wavelength units so we convert to frequency with:

$$\sigma_{\nu} = \sigma_{\lambda} \left(\frac{c}{\lambda_0^2}\right),$$
where $\lambda_0$ is the line rest wavelength. We verified our line fluxes by also computing the line flux via numerical integration and found both methods produced very similar results.

In the following sub-sections, we describe in more detail the features of the detected species.

\subsubsection{Atomic Hydrogen}
\begin{figure*}
  \centering
  \includegraphics[width=\textwidth, trim={2cm 0 1cm 0}, clip]{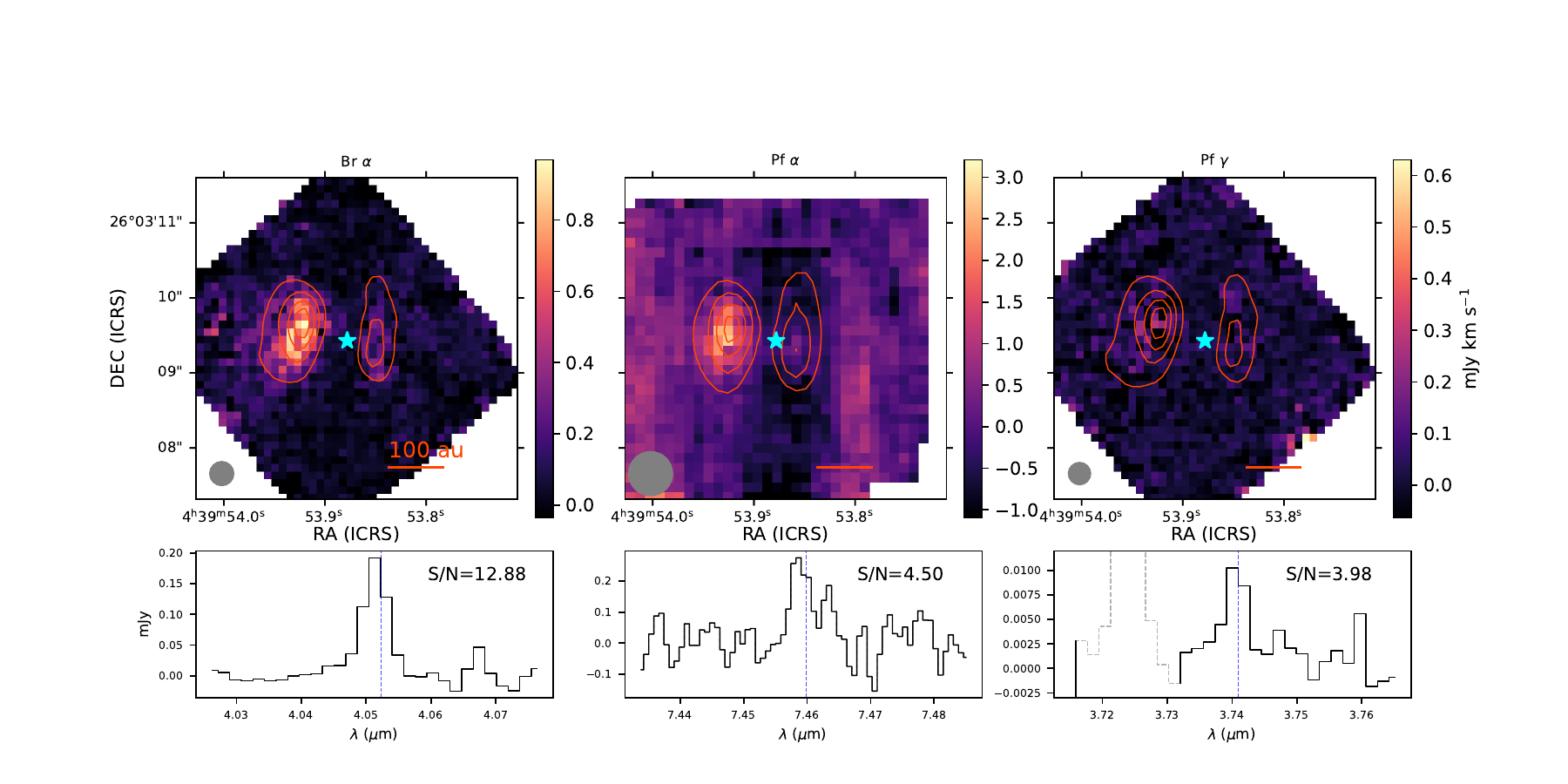}
  \caption{\textbf{Top row:} Continuum subtracted moment 0 maps for the three detected atomic hydrogen lines. The contours correspond to the local continuum emission. \textbf{Bottom row:} the corresponding line profiles for the atomic hydrogen lines along with the S/R of the line. The line profiles were extracted using the elliptical apertures shown in Figure \ref{fig:full_spec}. For Pf$\alpha$, only the eastern aperture was used as the western aperture introduced more noise than signal and thus made the line profile less clear. The gray dotted portion of the Pf $\gamma$ line profile is a nearby H$_2$ line that has been masked out.}
  \label{fig:h lines}
\end{figure*}

The atomic hydrogen lines are essential for this study because they can trace magnetospheric accretion activity. The three atomic hydrogen lines that we detect in our spectra are Br$\alpha$ and Pf$\gamma$ in NIRSpec, and Pf$\alpha$ in MIRI. Humphrey's $\beta$ also falls within our spectral coverage but is not detected. Moment 0 maps and line profiles for our detected atomic hydrogen species are shown in Figure \ref{fig:h lines}. 

The Br$\alpha$ emission appears strongly co-spatial with the continuum emission. In addition, it shows a brightness asymmetry between the eastern and western disk surfaces in the moment 0 map. Within the eastern disk surface, the Br$\alpha$ emission matches well with the continuum, shown as contours in Figure \ref{fig:h lines}. On the other hand, the emission from the western side is quite weak and is barely detected.

\begin{figure}
    \centering
    \includegraphics[width=\linewidth]{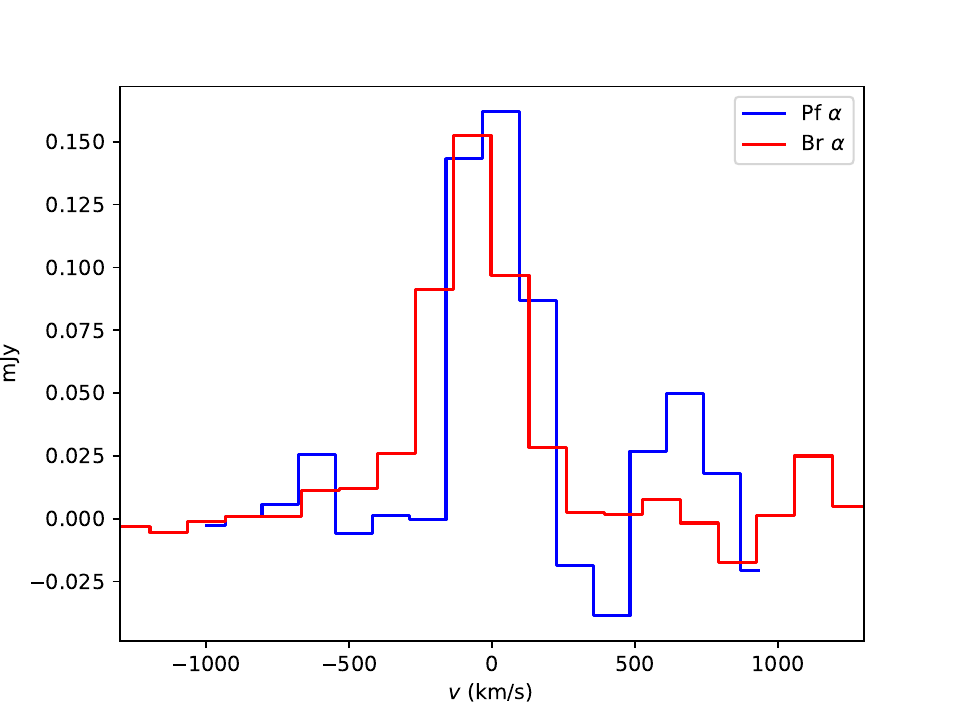}
    \caption{Comparison of the Pf$\alpha$ and Br$\alpha$ line profiles after the Pf$\alpha$ profile has been down-sampled to match the NIRSpec spectral resolution.}
    \label{fig:pfa rebinned}
\end{figure}

The Pf$\alpha$ emission shows a similar morphology to Br$\alpha$ but with significantly lower S/N. In addition, the east-west asymmetry is stronger in Pf$\alpha$ as there is no significant emission on the western side. In the line profile for Pf$\alpha$ it also appears that the line could be double peaked. To determine if line profile for Pf$\alpha$ is consistent with Br$\alpha$, we convolved the spectrum to match the spectral resolution of NIRSpec. We first convolved the line profile with a Gaussian kernel with a width given by

$$\sigma = \frac{\sqrt{\sigma_{NIRSpec}^2 - \sigma_{MIRI}^2}}{2\sqrt{2\ln2}}$$

enabling us to match the spectral resolution; $\sigma_{NIRSpec}$ and $\sigma_{MIRI}$ are the spectral resolutions of NIRSpec and MIRI respectively. We then re-binned the convolved line profile by a factor of 4 to match the velocity channel spacing of NIRSpec. A comparison of the Br$\alpha$ line profile and the convolved and regridded Pf$\alpha$ profile is shown in Figure \ref{fig:pfa rebinned}. We observe that once the spectral resolutions and sampling are matched, the two line profiles are indeed consistent.

The Pf$\gamma$ emission is the weakest of the atomic hydrogen lines. However, its emission morphology is consistent with the other lines, with the emission being concentrated more strongly on the eastern side.

\subsubsection{Molecular Hydrogen}
\begin{figure*}
    \centering
    \includegraphics[width=\linewidth, trim={2cm 0 1cm 0}, clip]{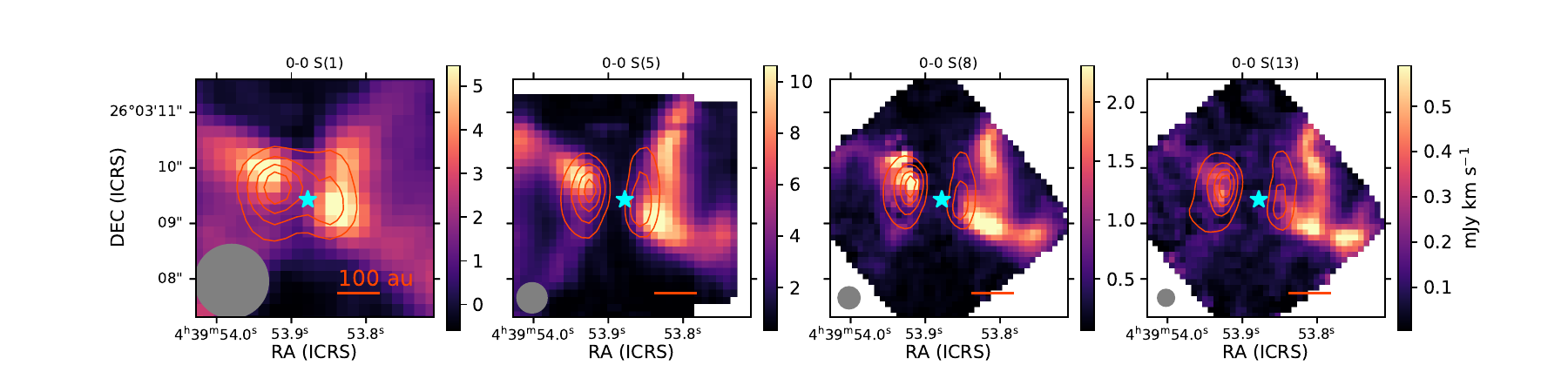}
    \caption{Continuum subtracted moment 0 maps for a sample of our detected molecular hydrogen transitions with contours representing the local continuum.}
    \label{fig: H2 grid}
\end{figure*}

We detect several rotational molecular hydrogen transitions ranging from H$_2$ 0-0 S(1) to H$_2$ 0-0 S(22). We show moment 0 maps for a subset of these transitions in Figure \ref{fig: H2 grid}. The emission for all of the H$_2$ transitions show a similar morphology. They all show arcs opening outward along the east-west direction originating near the center of the continuum emission for each side of the disk. These structures are consistent with other JWST studies that show H$_2$ tracing both the protostellar inner outflow cavity walls as well as the material inside the outflows themselves. \citep[e.g.][]{delabrosse2024, tychoniec2024}. 

The moment 0 maps for the longer wavelength transitions covered by MIRI (S(1) and S(5) in Figure \ref{fig: H2 grid}) show a point symmetry around the center of the continuum emission. The eastern outflow cavity shows a bright peak coinciding with the northern part of the continuum while the western outflow cavity shows a bright peak coinciding with the southern part of the continuum. Additionally, there is a strong decrease in emission in the southern half of the eastern outflow cavity, which is most clearly seen in the S(5) map.

The emission for the shorter wavelength transitions covered by NIRSpec (S(8) and S(13) in Figure \ref{fig: H2 grid}) show a stronger asymmetry between the eastern and western outflow cavities. For these transitions, the H$_2$ emission from the western outflow cavity is significantly stronger than the eastern side, which is most clearly shown in the S(13) map. This brighter outflow cavity exhibits a more complex morphology that is not apparent in the longer wavelength transitions, possibly due to a combination of angular resolution and excitation.

\subsubsection{Ions}
\begin{figure*}
    \centering
    \includegraphics[width=\linewidth, trim={2cm 0 1cm 0}, clip]{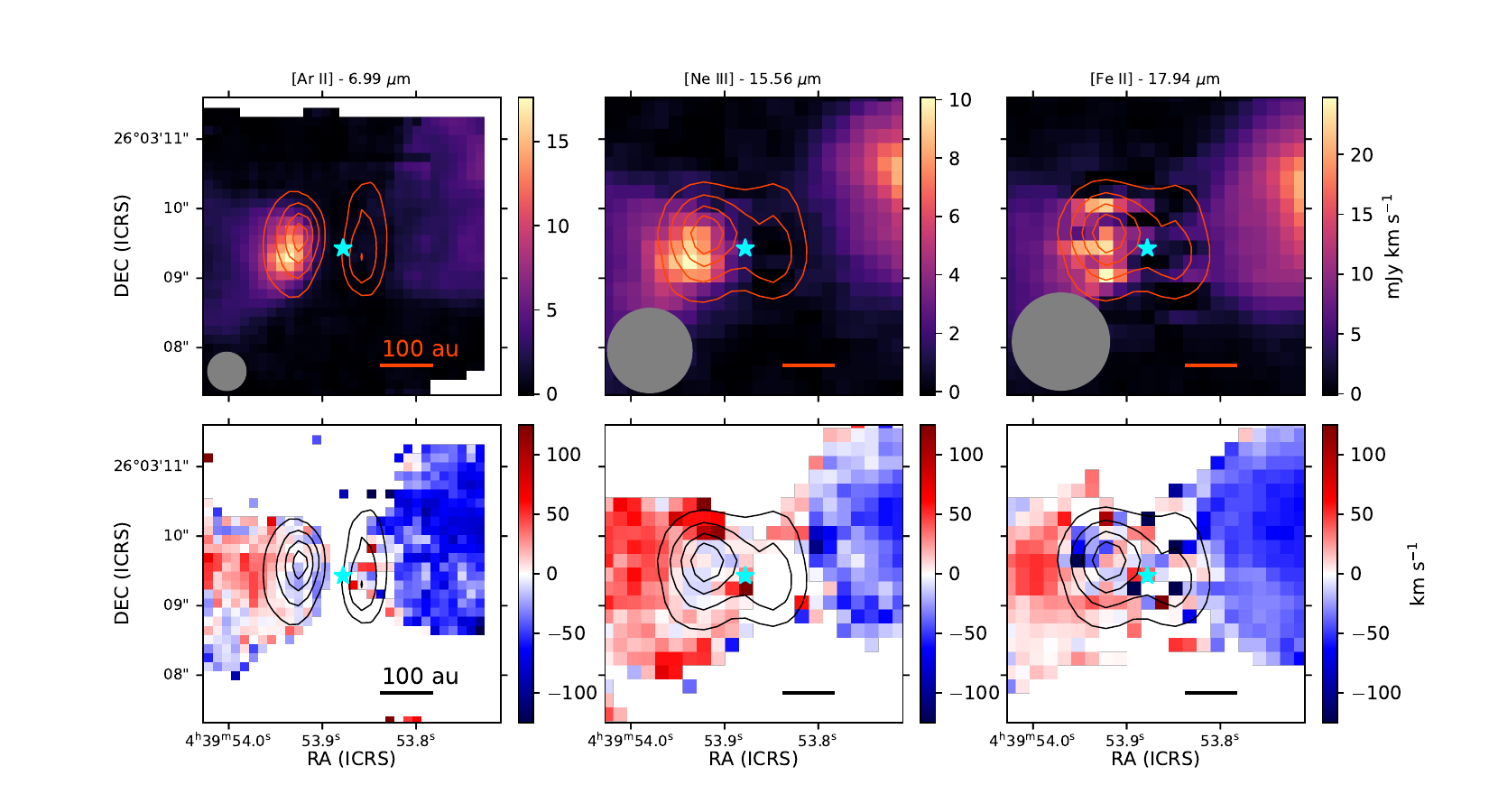}
    \caption{\textbf{Top row:} Continuum subtracted moment 0 maps for [Ar II], [Ne III], and [Fe II]. \textbf{Bottom row:} Continuum subtracted moment 1 maps for the same species. Pixels below the 3$\sigma$ level in the moment 0 maps have been masked out of the moment 1 maps.
 For both rows, the contours show the local continuum.}
    \label{fig: Ion grid}
\end{figure*}

We detect several atomic ions in our data, displaying various emission morphologies. We show the moment 0 and moment 1 maps for [Ar II], [Ne III], and [Fe II] in Figure \ref{fig: Ion grid}. The maps for [Ne III] and [Fe II] show a similar structure consisting of a narrow, outflow-like arc emerging from the eastern side of the continuum, and a second concentration of emission located west of the western side of the continuum. Although it appears outflow-like, the eastern emission structure has a narrower opening angle than is seen in the outflow cavities traced by H$_2$. This different morphology makes sense as [NeIII] traces the ionized atomic cavity while H$_2$ traces low-velocity shocks along the cavity wall. The western emission structure, though less arc-like, is of comparable width. Additionally, the axis defined by these two emission structures is misaligned to the axis perpendicular to the disk. The eastern structure is located south of the disk, while the western structure is north of the disk. In [Ar II] we see a similar morphology to [Ne III] and [Fe II], although the western emission structure is not present. However, this is potentially due to the smaller FOV of the [Ar II] map. Although the bright emission is not present, there is weak, cone-shaped structure emerging from the western side. Additionally, the higher spatial resolution of the [Ar II] map enables a clearer view of the eastern, arc-like structure.

We also show moment 1 maps for these ions in Figure \ref{fig: Ion grid}, to examine the velocity structure of the emission. For all three ions shown in Figure \ref{fig: Ion grid}, we see a clear velocity gradient with eastern emission being red shifted and the western emission being blue shifted. In the velocity map for [Ar II], we see an additional blue-shifted velocity component on the eastern edge of the red-shifted emission. For these ions, we see average blue/red-shifted velocities of $\sim20-40$ km s$^{-1}$.

The red and blue shifted sides of the jet emission shown in these moment 1 maps are swapped compared to what is observed for the larger scale swept-up outflow gas seen in millimeter CO data \citep{bontemps1996, hogerheijde1998}. However, this orientation is consistent with observations of SiO presented in \citet{van'thoff2023} and [NeII] from Devaraj et. al. (Submitted).

For the MIRI H$_2$ and atomic lines, we also find similar emission morphologies as were shown in Devaraj et al. (Submitted), however, our data has better spatial coverage of the western extended emission.

\subsection{Water and OH}
\begin{figure*}
    \centering
    \includegraphics[scale=0.65]{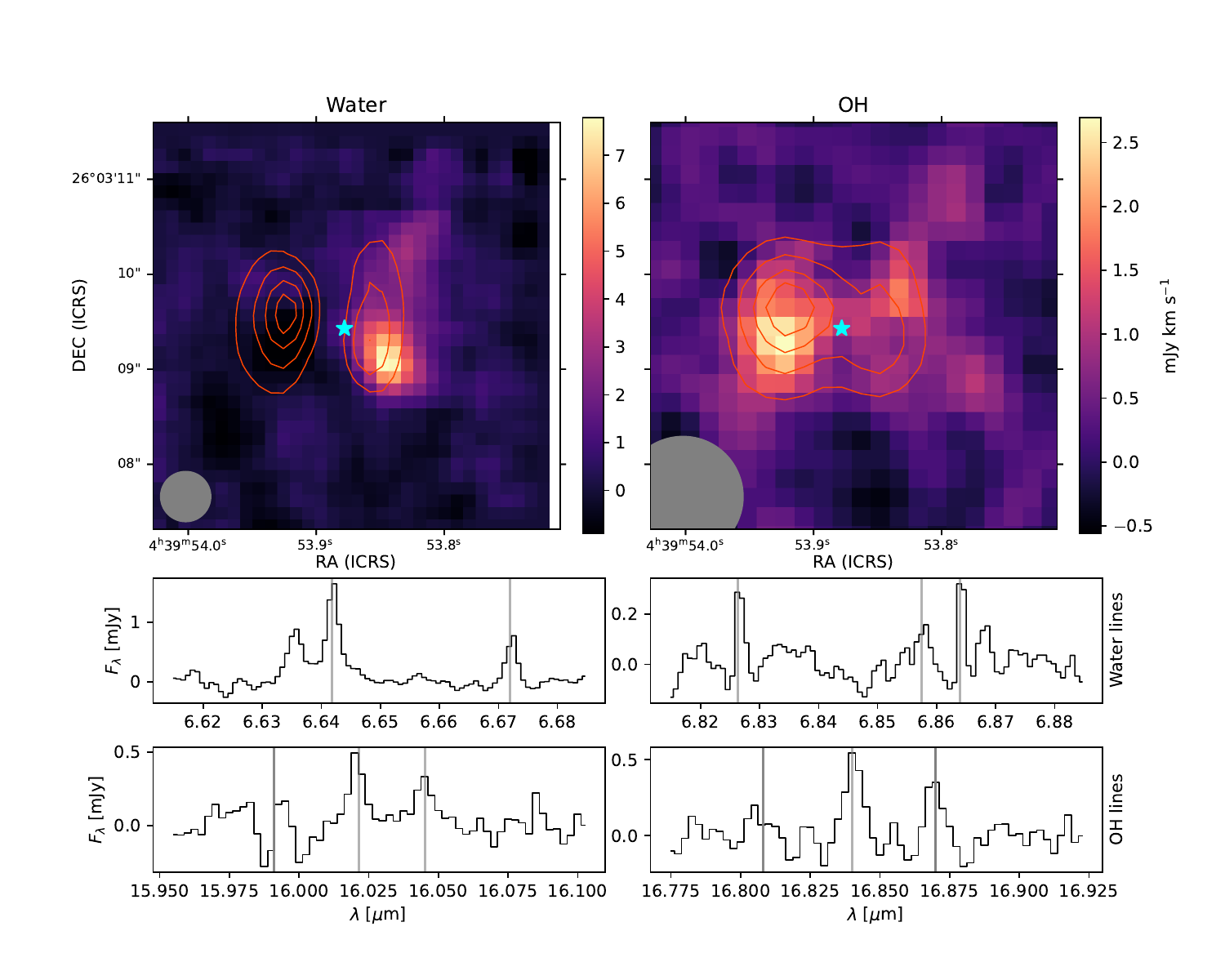}
    \caption{Stacked moment 0 maps for water (left) and OH (right). The contours represent the 16.0 $\mu$m continuum (for OH), and the 6.6 $\mu$m continuum (for water). The top row of spectra show the detected water lines and the bottom row of spectra show the detected OH lines.}
    \label{fig: water and oh}
\end{figure*}

We additionally detect several water and OH transitions in our spectrum \citep[see also][]{vanGelder2024}. The emission from these species is quite weak, so we stacked together the moment 0 maps for all of the water transitions and all of the OH transitions to create a single, higher S/N map for each species, shown in Figure \ref{fig: water and oh}.

Similar to the atomic hydrogen lines, the OH emission is concentrated on the eastern side of the continuum. However, due to the noise and resolution it is difficult to determine the morphology of the OH, but it is broadly similar to Br$\alpha$ and Pf$\alpha$. 
Contrastingly, the water emission is located only on the western side of the continuum. Thus we observe that the peaks of the OH and water emission appear to be anti-correlated. The morphology of the water emission appears similar to the western outflow cavity traced by H$_2$, but more truncated in the north-south direction. Like the H$_2$, the water emission is stronger towards the south and weaker towards the north.

\section{Analysis} \label{sec: analysis}
In this section we will analyze the Br$\alpha$ emission to first verify that its emission is consistent with tracing accretion. We will then compute an intrinsic Br$\alpha$ luminosity in order to estimate the accretion luminosity, $L_{acc}$ of L1527. 

\subsection{Line to Continuum Ratio Maps}
\begin{figure*}
    \centering
    \includegraphics[width=\linewidth]{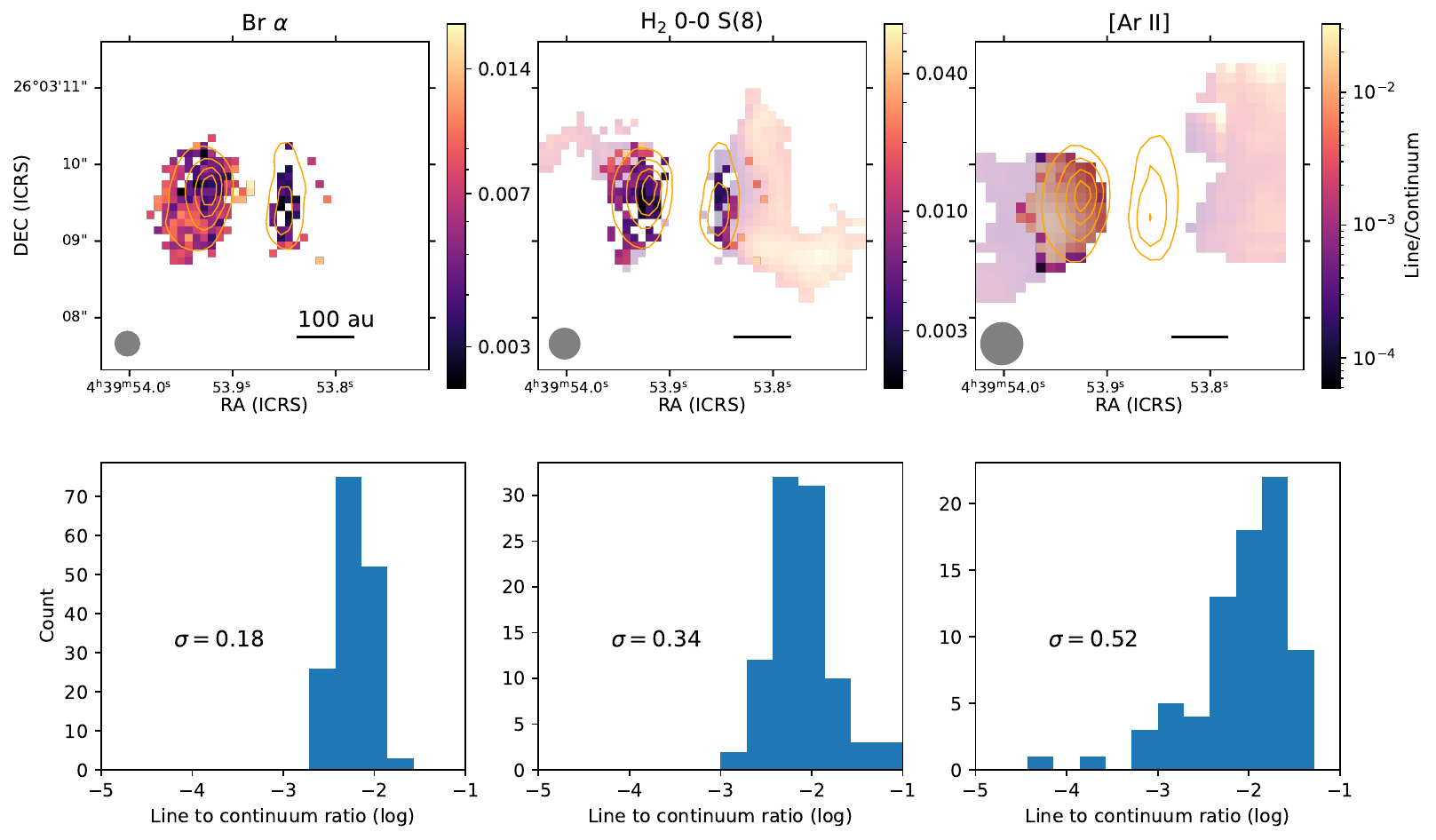}
    \caption{\textbf{Top:} Line to continuum ratio maps for Br$\alpha$, H$_2$ 0-0 S(8), and [Ar II]. We have masked out the regions where the line emission is below the 1 $\sigma$ threshold. Additionally, we emphasize the inner region of the map where the Br~$\alpha$ emission is present by making regions outside of this fainter. The contours represent the local continuum. \textbf{Bottom}: Histograms of the line to continuum ratio maps along with the standard deviation. These histograms are generated from only the compact emission, where Br~$\alpha$ is present, to better compare the different species on the same spatial scales. These maps and histograms show that the line to continuum ratio is more constant for Br$\alpha$ compared to H$_2$ and [Ar II], supporting the interpretation that the Br$\alpha$ is likely being observed in scattered light.}
    \label{fig:ratio maps}
\end{figure*}

It is important to verify that the atomic hydrogen lines are indeed tracing accretion and are not instead being locally produced from other processes such as shocks in jets, as seen in other sources \citep[e.g,][]{Harsono2023, Federman2024}. The continuum at 5 $\mu$m is dominated by emission from the inner disk walls, just outside of the dust sublimation radius, located within 1 au of the protostar \citep{natta2001, mcclure2013}. Thus, if the hydrogen lines are from accretion, both the continuum and the Br$\alpha$ would be produced in the obscured inner disk region and be detected in scattered light. 
% To confirm this is the case, we created line to continuum ratio maps for Br$\alpha$ along with H$_2$ and [Fe II], shown Figure \ref{fig:ratio maps}. If the emission is being observed in scattered light, we would expect the ratio map to be spatially constant, since the line and continuum emission should have the same structure. These maps simply show the ratio between the moment 0 maps for the continuum-subtracted line cubes and the local continuum over the same wavelengths. We use the Br$\alpha$ line for atomic hydrogen since it has the strongest emission among our atomic hydrogen lines. We also expect stronger scattering of continuum and line photons at this wavelength since scattering efficiency drops with increasing wavelength. We use H$_2$ and [Fe II] as comparison since they both show extended emission that is locally produced by shocks resulting from the outflow.
To confirm this is the case, we created line to continuum ratio maps for Br~$\alpha$ along with H$_2$ S(8) and [Ar II], shown in Figure \ref{fig:ratio maps}. If both the continuum and Br~$\alpha$ emission originated from the star or inner disk, we would expect the ratio map to appear spatially constant since the line and continuum emission should have the same structure. These maps simply show the ratio between the moment 0 maps for the continuum-subtracted line cubes and the local continuum over the same wavelengths. We use the Br$\alpha$ line for atomic hydrogen since it has the strongest emission among our atomic hydrogen lines. We also expect stronger scattering of continuum and line photons at this wavelength since scattering efficiency drops with increasing wavelength. We use H$_2$ S(8) and [Ar II] as comparison since they have the highest spatial resolution among our observed species that trace extended emission locally produced by shocks resulting from the outflow. In Figure \ref{fig:ratio maps}, we also highlight the spatial regions where Br~$\alpha$ emission is located in the H$_2$ and [Ar II] maps. To isolate this region, we determine a mask from the Br~$\alpha$ moment 0 map based on where the flux is above the 1 $\sigma$ threshold. We then apply this mask to the other 2 species, using the \texttt{reproject} to apply the mask to the [Ar II] map where the pixel sizes differ from Br~$\alpha$.

% These maps indicate that atomic hydrogen is likely being observed in scattered light. First, the emission from Br$\alpha$ is spatially concentrated within the scattered light continuum. Additionally, the line to continuum ratio is relatively spatially constant, which implies that the emission is mixed with the continuum. In contrast, the line to continuum ratio maps for H$_2$ and [Fe II] are clearly not spatially coincident with the continuum, and show strong spatial variations. This is consistent with these lines being locally produced by shock heating. Additionally, the histograms in Figure \ref{fig:ratio maps} demonstrate that the line to continuum ratio for Br$\alpha$ is much more constant than H$_2$ or [Fe II]. The standard deviations shown in the histograms show that the spread of the log of the line to continuum ratio is a factor of $\sim2-4$ times smaller for Br$\alpha$ compared to H$_2$ and [Fe II]. This implies that the line to continuum ratio for H$_2$ and [Fe II] cover several more orders of magnitude than Br$\alpha$.

These maps indicate that atomic hydrogen is likely being observed in scattered light. First, the emission from Br~$\alpha$ is spatially concentrated within the scattered light continuum. Additionally, the line to continuum ratio is relatively spatially constant, which implies that the emission is mixed with the continuum. In contrast, the line to continuum ratio maps for H$_2$ and [Ar II] show extended emission beyond the extent of the continuum. This is consistent with these lines being locally produced by shock heating. Additionally, the histograms in Figure \ref{fig:ratio maps} demonstrate that the line to continuum ratio for Br$\alpha$ is much more constant than H$_2$ or [Ar II]. The standard deviations shown in the histograms show that the spread of the log of the line to continuum ratio is a factor of $\sim2-3$ times smaller for Br~$\alpha$ compared to H$_2$ and [Ar II]. This implies that the line to continuum ratio for H$_2$ and [Ar II] is spread out several more order of magnitude than Br$\alpha$. However, we cannot fully rule out that some of the Br~$\alpha$ emission could be produced by shock heating, especially if the Br~$\alpha$ emission is faint enough that extended jet emission is below our sensitivity level. We do not believe this would be a major contributor to the Br~$\alpha$ emission because other low-mass sources that have collimated jet emission in [Fe~II], [Ar II], etc., also have strong Br$\alpha$ jet emission that is co-spatial with those other jet tracers \citep[e.g.][]{Federman2024}, which we do not observe.

\subsection{Accretion Luminosity}
Since the line to continuum maps indicate that atomic hydrogen traces accretion in scattered light, we can use it to determine the accretion luminosity of L1527. We use the empirical relation from \citet{Komarova&Fischer2020} to relate Br$\alpha$ luminosity to accretion luminosity:

\begin{equation}\label{eq: K and F}
    \log(L_{\text{acc}} / L_\odot) = (1.81 \pm 0.11)\log(L_{\text{Br}\alpha}/L_\odot) + (6.45 \pm 0.38)
\end{equation}

To compute the Br$\alpha$ luminosity, we must determine how much Br$\alpha$ flux we are missing due to observing it in scattered light and due to extinction from the disk and through the envelope. Since the Br$\alpha$ emission is being produced primarily between the inner disk and the protostar, we have to consider how much flux is getting scattered by the disk surface out of our line of sight or is obscured by the disk. We then determine how much of the remaining flux gets extincted by the surrounding envelope. Once those effects are corrected for, we can simply use the distance to L1527 to calculate the Br$\alpha$ luminosity from the corrected line flux. The issue of observing accretion lines through scattering has already shown that when these effects are not accounted for, the estimated accretion luminosities are extremely low \citep{delabrosse2024}.

\subsubsection{Scattering}\label{sec:scattering}

To estimate the flux loss due to scattering out of our line of sight and extinction by the edge-on disk, we generated radiative transfer (RT) models of L1527 IRS protostellar system.

Our goal is to characterize to what extent the scattered emission from the accretion shocks on the protostellar surface and the accretion funnels from the inner disk to the protostar could be detected in scattered light. We use the \texttt{hyperion} radiative transfer code \citep{robitaille2011} to estimate how much of the intrinsic flux we are actually able to observe in scattered light. We first set up a model consisting of the protostar, a disk, and a an envelope. We use the flared disk, a rotationally flattened envelope \citep{ulrich1976}, and bipolar cavity models included in the \texttt{hyperion} code \citep{robitaille2011}. To create our model, we used the stellar, disk, and envelope parameters derived by \citet{tobin2013} using SED fitting. We describe of our model parameters in Appendix \ref{appendix: RT}. We generated models with various disk masses ranging from 0.001 to 0.5 M$_\odot$ since the disk mass is not very well constrained and has the largest impact on the amount of light blocked by the disk. There is some degeneracy with altering the disk mass or the scale height, as both of these parameters will impact the width of the dark absorption lane, which is key in setting the amount of solid angle that is extincted by the disk. Since our goal is not to constrain the physical disk properties and we are instead only interested in the amount of scattered light we observe, it is acceptable that we do not break this degeneracy, thus we keep the scale height parameters fixed for simplicity. For the dust properties, we used the opacity laws and scattering phase function from the KP5 opacity law \citep{pontoppidan2024}. The KP5 model fits the 1-1000 $\mu$m extinction curve of dense molecular clouds and protostellar envelopes from Spitzer and JCMT observations. KP5 incorporates both graphite and amorphous silicate dust populations with icy mantles. To generate our synthetic images, we ran a monochromatic, scattering only simulation at 4.05 $\mu$m (corresponding to Br$\alpha$), then convolved the result with the NIRSpec PSF obtained using the \texttt{stpsf} package \citep{perrin2014}.

\begin{figure}
    \centering
    \includegraphics[width=\linewidth]{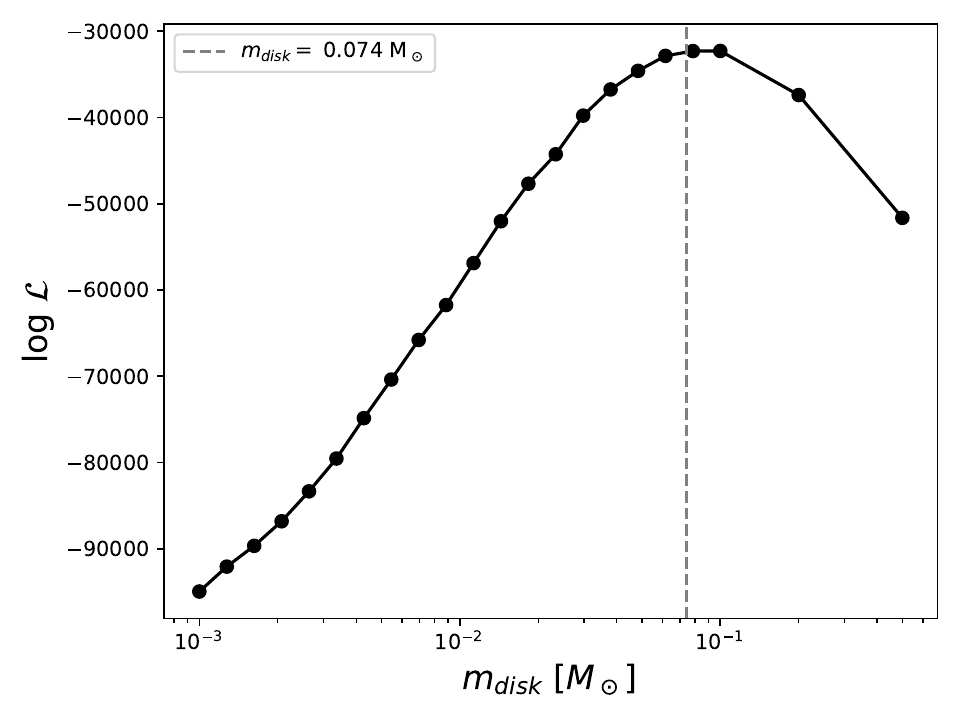}
    \caption{The log-likelihoods for our RT models as a function of disk mass. The vertical line corresponds to our best model which was determined by using the log-likelihoods to compute a weighted average of the disk masses.}
    \label{fig:mdisk v L}
\end{figure}

To determine the best disk mass to use for our model, we compared each synthetic image to the real NIRSpec 4 $\mu$m image and computed a log-likelihood, shown in Figure \ref{fig:mdisk v L}. We estimated the uncertainty on our NIRSpec data by taking the standard deviation of the image after masking out regions with signal $> 3\sigma$, then computed the log-likelihood via the Chi-squared. We ensured that both the real and synthetic data were in units of Jy, so that the fluxes could be appropriately compared.

We calculated our best disk mass by using the log-likelihoods to compute a weighted average of the disk masses and found the best value to be $m_{disk} = 0.074$~$M_\odot$, which is shown as a dashed vertical line in Figure \ref{fig:mdisk v L}.

This value is a few times larger than the disk mass of $\sim0.01-0.03~\Msun$ derived using the mm flux density \citep{van'thoff2023}. This is not a big concern since, for the purpose of this work, we are more focused on producing a model that most closely matches the data, rather than getting the best constraints of the physical parameters of the system. A more thorough study using an MCMC or nested sampling approach and many more free model parameters would be needed to best characterize the disk and envelope parameters, but this is beyond the scope of this work.

\begin{figure*}
    \centering
    \includegraphics[width=\linewidth, trim={0 1cm 0 1cm}, clip]{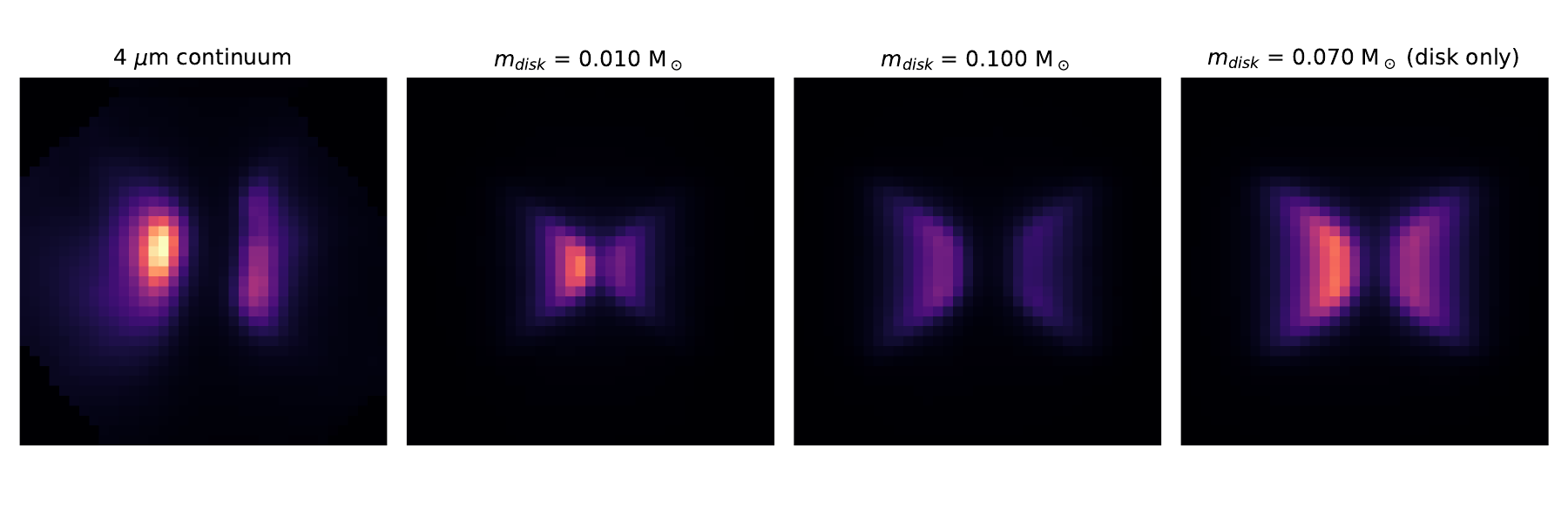}
    \caption{The 4 micron NIRSpec continuum image compared with RT models with various disk masses. The RT models have the same pixel size as the 4 micron image and have been convolved with the NIRSpec PSF. All four images are shown on the same color-scale set by the real 4 $\mu$m continuum image. The middle two RT models include both a disk and an envelope. Their morphologies and fluxes indicate that we expect the disk mass to be between 0.01 and  $0.1 ~M_{\odot}$. The right most model only includes the disk and is what we used to calculate the fraction of intrinsic flux were are able to observe in scattered light. We show all of our disk + envelope RT models in Appendix \ref{appendix: RT}.}
    \label{fig:rt model examples}
\end{figure*}

To then determine the fraction of intrinsic Br$\alpha$ emission that gets through the disk (i.e is not blocked by the disk and is scattered into our line of sight) we used our derived disk mass and computed a RT model in the same way as before, but now only included the protostar and the disk. We then compared the total flux of this model to what we would expect from an un-obstructed star. Here we are assuming that the light from accretion is scattered in the same way as the protostellar light since it originates very close by to the embedded protostar. Additionally, we assume that most of the light we observe originates from scattering off the disk rather than the inner envelope. To confirm this, we numerically integrate the envelope density profile to obtain an optical depth $\tau$. We attenuated the total flux of the disk-only model by this $\tau$ and obtain a flux of 0.0016 Jy. This value is very similar to the total flux we observe in disk+envelope model of 0.0020 Jy, implying that most of the flux is coming from the disk. We calculated the specific luminosity at the Br$\alpha$ frequency via the equation for monochromatic luminosity given by:
$$L_\nu = 4\pi^2R^2B _\nu(T),$$ where $R$ is the stellar radius and $B_\nu$ is the Planck function which we evaluate at $T=4000 K$. We increased the stellar radius, $R$, to 3.45 R$_\odot$ when computing $L_\nu$, so that the total luminosity agreed with observations. If we instead used the true estimated value for stellar radius (3 R$_\odot$), we would underestimate the luminosity, as we would only be considering the photospheric luminosity and not the accretion luminosity. We then use the distance to the source of 140 pc to convert this to the specific flux, $F_{\nu}$, we would expect to observe if there was no obscuring disk or envelope. To then calculate the fraction of intrinsic Br$\alpha$ flux that emerges through the disk, $f_{scat}$, we calculate the total flux of the disk only RT model, and divide by $F_{\nu}$. To estimate the uncertainty, we computed the standard deviation of the $f_{scat}$ values corresponding to the 10 disk masses with the highest likelihood. This results in a quite small scattering fraction of $f_{scat} = 0.008 \pm 0.001$. Since this is purely a statistical uncertainty, there are likely systematic uncertainties from our analytic disk+envelope model that are not being included.

Thus, we can estimate to a factor of a few, the amount of light from the central protostar (or accretion from the inner disk to protostar) that should be transmitted by the disk via scattering. While we could include the envelope in the model to estimate how much the scattered light is further attenuated, the non-spherical envelope density structure and uncertain radius compels us to turn to an empirical estimate of the extinction through the envelope.

\subsubsection{Extinction}
\begin{figure}
    \centering
    \includegraphics[width=\linewidth]{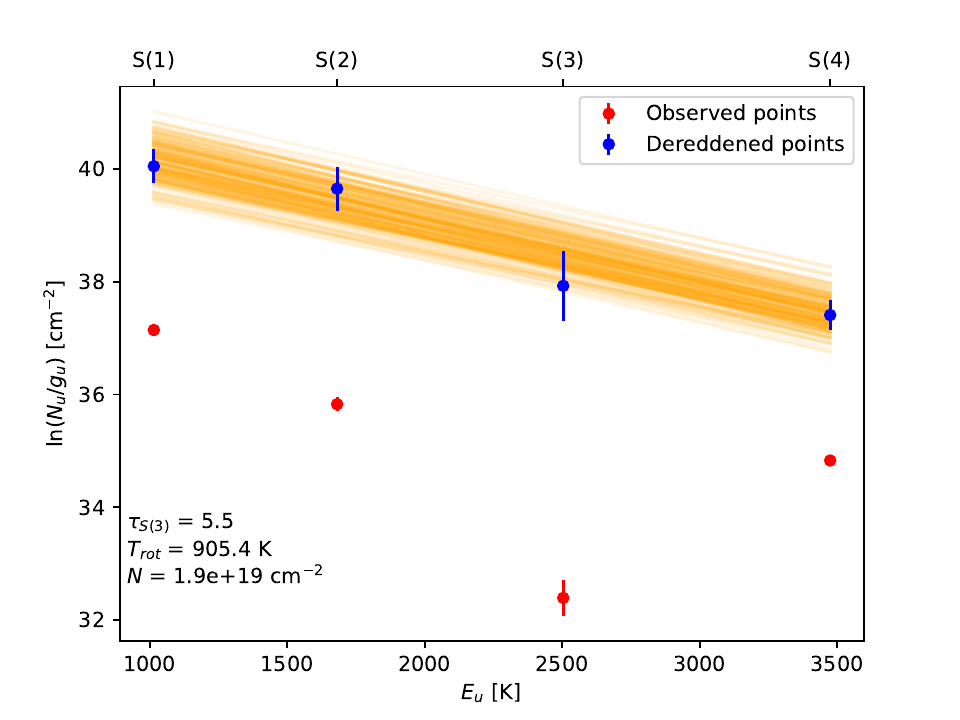}
    \caption{Rotation diagram before and after de-reddening based on the extinction fit. The orange lines show 200 samples from the MCMC fitting. We show the resulting posteriors from this fit in a corner plot in Appendix \ref{appendix: extinction}.}
    \label{fig:rot curve}
\end{figure}

We determined the extinction through the envelope by fitting the H$_2$ rotation diagram \citep{narang2024, francis2025}. We used the H$_2$ 0-0 S(1)-S(4) lines to generate the rotation diagram which is shown in Figure \ref{fig:rot curve}.

The x-axis of the rotation diagram shows $E_u$, the upper state energy for the transition. The y-axis shows $\ln(N_u/g_u)$ where $N_u$ is the column density of H$_2$ molecules in the upper energy state and $g_u$ is the degeneracy of this state. We compute $N_u$ using the relation:
$$N_u = \frac{4 \pi I_{u,l}}{A_{u,l}h\nu}$$
where $I_{u,l}$ is the line intensity, $\nu_{u,l}$ is the frequency of the transition, $h$ is Planck's constant, $c$ is the speed of light, and $A_{u, l}$ is the Einstein A coefficient. We converted from line flux to intensity using the area in steradians of the apertures we used to extract the line profiles.

We did not have the spectral coverage to observe all of the H$_2$ 0-0 S(1)-S(4) lines, so we used data from the JOYS program presented in  Devaraj et al. (Submitted) for the lines that fell in a short or medium MIRI sub-band. To calculate the H$_2$ line fluxes, we used the elliptical apertures shown in Figure \ref{fig:full_spec} to consider the H$_2$ emission that is spatially coincident with the Br$\alpha$. These line profiles and fits are shown in Appendix \ref{appendix: extinction}. We can use the rotation diagram to simultaneously fit for the H$_2$ column density ($N$), temperature ($T$), and extinction at the H$_2$ 0-0 S(3) wavelength of 9.3 $\mu$m ($\tau_{S(3)})$ via MCMC using the relation:

$$\ln\left(\frac{N_u}{g_u}\right) = \ln\left(N\frac{\exp(-E_u/T)}{Q(T)}e^{-\tau_\lambda}\right),$$

where we compute the wavelength-dependent $\tau_\lambda$ via:

$$\tau_\lambda = \tau_{S(3)} \left(\frac{\kappa_\lambda}{\kappa_{S(3)}}\right),$$

where $\kappa_\lambda$ is the wavelength dependent opacity where we again use the KP5 opacity law \citep{pontoppidan2024}, and $\kappa_{S(3)}$ is this opacity evaluated at 9.3$\mu$m.

In Figure \ref{fig:rot curve}, we show the results of the fitting by showing the original data points in red and the de-reddened points in blue. The blue points now appear strongly linear, which indicates that we have corrected for extinction well. We use our best fit $\tau_{S(3)}$ and the KP5 opacity law to calculate the optical depth for Br$\alpha$ (4.05 $\mu$m) due to extinction to be $\tau_{ext, Br\alpha} = 2.52 \pm 0.24$.

\subsubsection{Accretion Luminosity}
We can now estimate the scattering and extinction corrected Br$\alpha$ flux via:

$$F_{Br\alpha} = F_{Br\alpha, obs}\frac{1}{f_{scat}}e^{\tau_{Br\alpha}} = 2.50 \pm 0.69\times10^{-13} \ \text{erg }\text{s}^{-1}\text{ cm}^{-2},$$

where we computed $F_{Br\alpha, obs}$ using the elliptical apertures shown in Figure \ref{fig:full_spec}.

Using a distance of 140 pc to L1527, this implies a Br$\alpha$ luminosity of $L_{Br\alpha} = 1.53 \pm 0.43 \times10^{-4} \ L_\odot$ and an accretion luminosity of $L_{acc} = 0.35^{+0.16}_{-0.13}~L_\odot$ uisng equation \ref{eq: K and F}. We obtained the uncertainty on $L_{Br\alpha}$ through error propagation, where the error on $\tau_{Br\alpha}$ was obtained from the posterior of the MCMC fitting. We used the uncertainties from \citet{Komarova&Fischer2020} to determine the upper and lower uncertainties on $L_{acc}$.

\section{Discussion} \label{sec:discussion}
In this section, we will discuss the accretion mechanism for L1527 IRS and estimate the accretion rate. We will also discuss potential evidence in the data that supports a non-steady and asymmetric accretion scenario.

\subsection{Mode of Accretion}
The detection of several atomic hydrogen lines consistent with being viewed in scattered light indicates that accretion is likely occurring via the magnetosphere. Accretion via a boundary layer would not produce such bright HI emission since the warm region producing the emission is confined to a much smaller area compared to the magnetosphere. This is consistent with other work investigating protostellar accretion with NIR spectroscopy where they also find evidence for magnetospheric accretion \citep[e.g.][]{laos2021}. However, other studies have argued that Class 0s maybe undergoing a different accretion mechanism than Class Is in scenarios with stronger accretion and a lack of detected atomic hydrogen lines \citep[e.g.][]{leGouellec2024, leGouellec2025}.

\subsection{Mass Accretion Rate}

The accretion rate of L1527 can be estimated from our calculated accretion luminosity using the approximation \citep{calvet_gullbring1998}:

$$L_{acc} = 0.8\frac{GM_*\dot{M}}{R_*} \Rightarrow \dot{M} = 1.25\frac{L_{acc}R_*}{GM_*}.$$

$L_{acc}$ is the accretion luminosity, $G$ is the gravitational constant, $M_*$ is the protostellar mass, $\dot{M}$ is the accretion rate, and $R_*$ is the protostellar radius.

To estimate the protostellar mass, we will use the results from modeling the gas kinematics of ALMA data of L1527 that was taken as part of the eDisk large program: $M_*=0.40 \pm 0.06~M_\odot$ (Drechsler et al. in prep). Using the protostellar mass-radius relation from \citet{hartmann2025}, we adopt a protostellar radius of $R_*= 3~R_\odot$.  We calculate an accretion rate of

$$\dot{M} = 1.0 \pm 0.2 \times 10^{-7} \ M_\odot \text{yr}^{-1}$$

This accretion rate is too low to accrete the observed stellar mass in $\sim$ 0.1 or even 1 Myr. This discrepancy is evidence that the accretion rate could not have been constant throughout the life of the protostar. Specifically, this implies that the accretion rate must have been higher at some point in L1527's past in order to explain its current mass. This is consistent with theories that protostellar accretion is not constant and may be episodic or undergo accretion bursts where a large amount of material is accreted in a short amount of time \citep{kenyon1990, dunham2010, fischer2023}.

\citet{tobin2008} performed SED modeling and calculated a total system luminosity of $L_{tot} = 2.75~\Lsun$. Since our estimated accretion luminosity is only $L_{acc} \approx 0.4~\Lsun$, we find that the protostellar luminosity dominates the energy budget of the system. This is contradictory to some expectations for such a young source. As a result, this could imply that L1527 is older than we expect and could be classified incorrectly as a Class 0 source due its edge on geometry impeding the amount of observed stellar flux. However, it is estimated that the protostellar envelope around L1527 contains $\sim 1~\Msun$ of material within 0.05 pc \citep{chandler&richer2000, tobin2011}, which is greater than the current protostellar mass by a factor of $\sim 2$. Additionally, \citet{kristensen2012} estimated a similar envelope mass of $0.9~\Msun$ using radiative transfer models of both the SED and sub-mm spatial extent. Thus, L1527 would still be classified as Class 0 using the inclination-independent classifier $M_{env}/M_*$ \citep{andre1993}. 

\subsection{Accretion Asymmetry}
\begin{figure*}
    \centering
    \includegraphics[width=\linewidth, trim={2cm 0 2cm 0}, clip]{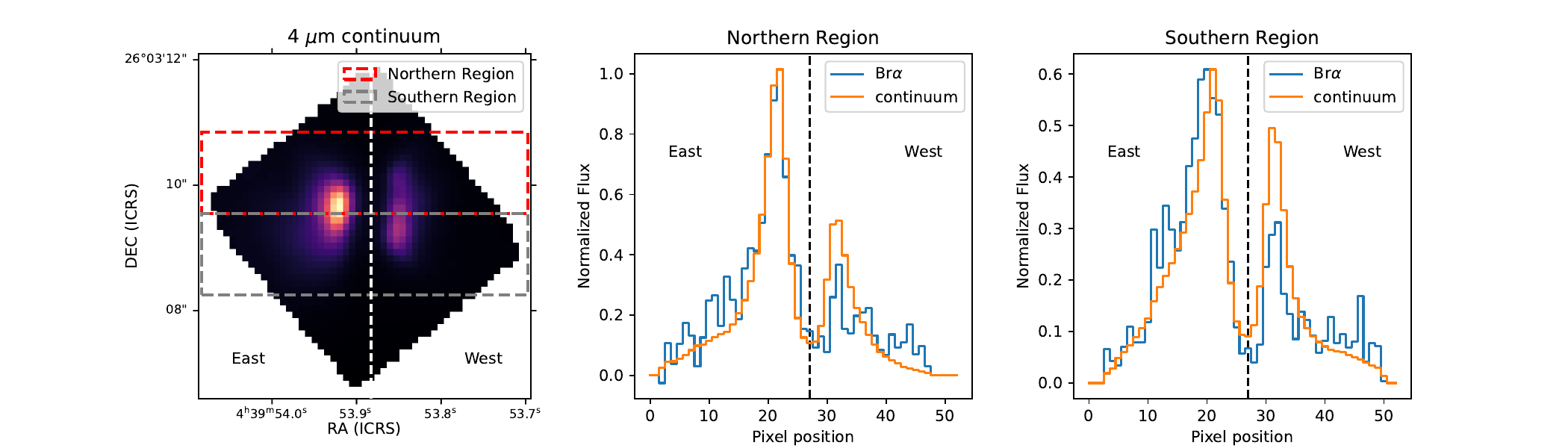}
    \caption{Horizontal profiles for the Br$\alpha$ (blue) and local continuum (red) moment 0 maps. Separate profiles are shown for the northern and southern regions of the emission maps. These profiles were created by summing the two regions shown by the dashed rectangles in the left plot along the vertical axis.}
    \label{fig:accretion asym NS}
\end{figure*}

Figure \ref{fig:h lines} shows stronger atomic hydrogen lines on the eastern disk surface and little to no emission on the western side. This could potentially imply an intrinsic asymmetry in the accretion, or it could be an inclination effect causing the western side to be more heavily extincted than the eastern side since it is slightly facing away from us. To distinguish between these two scenarios, we examined emission profiles of the Br$\alpha$ line and local continuum emission maps, shown in Figure \ref{fig:accretion asym NS}. If this asymmetry were an extinction effect, we would expect the flux to be reduced by a similar factor from the eastern to western sides for both the line and continuum emission. 

In Figure \ref{fig:accretion asym NS}, we see that in both the northern and southern region of L1527, the Br$\alpha$ line emission decreases by a larger fraction than the local continuum. This could imply an intrinsic asymmetry in the accretion, causing the Br$\alpha$ emission to be stronger on the eastern side. Since we assume that the line and continuum emission are being observed in scattered light, they should both experience the same amount of extinction. Thus, since their flux is not decreasing by the same amount, it is possible that there is less atomic hydrogen emission on the western side. This could imply that the material is not accreting onto the protostar isotropically, and is instead preferentially arriving on the eastern side.

\subsection{OH as a sign of accretion}
The most direct way to determine accretion rates and luminosities is to use the excess UV continuum flux \citep{calvet_gullbring1998} which is produced by the accretion shock onto the protostar. However, this emission is only detectable for more evolved pre-main sequence stars, not protostars which are still deeply embedded. Our detection of OH could indicate the existence of a UV field since it can be produced by the photo-dissociation of water by UV radiation \citep{tabone2021, watson2025}. The morphology of the OH and water emission shown in Figure \ref{fig: water and oh} support the explanation that the OH is due to the dissociation of water from UV accretion flux since their emission is spatially anti-correlated. Since the water is being photo-dissociated into OH in this scenario, it makes sense that where OH is present, there should not be much water and vice-versa. The emission morphology of the OH and water maps would then imply that the UV field is stronger on the eastern side (where the OH emission is stronger and the water emission is weaker). This is consistent with explanation that the trends seen in Figure \ref{fig:accretion asym NS} are a result of asymmetrical accretion that is stronger on the eastern side of the disk. 

However we need OH lines in the 9-10 $\mu$m range to verify that the OH is indeed due to photo-dissociation of water, and we lack spectral coverage over this range. The OH lines that we detect at around 16 $\mu$m can additionally be produced chemical pumping through the O + H$_2$ $\rightarrow$ OH reaction \citep{tabone2021}, which would be unrelated to accretion.

\section{Summary and Conclusions} \label{sec:conclusion}

We present NIRSpec and MIRI data toward the Class 0 protostar L1527 at 3-28 $\mu$m. We detect emission lines from atomic hydrogen, H$_2$, OH, water, and several ionic species. We studied these features and created emission maps, as well as used the accretion-tracing atomic hydrogen lines to learn about accretion in this source. Our key takeaways are summarized below: 

\begin{itemize}
    \item We detect the Br$\alpha$, Pf$\alpha$, and Pf $\gamma$ atomic hydrogen lines with the Br$\alpha$ emission being the strongest. All three of these lines appear co-spatial with the continuum emission and show significantly stronger emission on the eastern disk surface. The presence of these atomic hydrogen lines viewed in scattered light is an indication that accretion is occurring via the magnetosphere rather than a boundary layer.
    \item The molecular hydrogen emission traces wide angle, outflow cavities. While, the ionic species, such as [Ar II], [Ne III], and [Fe II], trace a more more narrow outflow structure with the western component of the emission beginning outside of the continuum. In addition, velocity maps for these ions reveal the western side as blue-shifted and the eastern as red-shifted, which is counter to what was expected based on previous molecular line data and the scattered light continuum.
    \item $L_{Br\alpha}$ was determined in order to calculate $L_{acc}$. We determined the intrinsic Br$\alpha$ flux by correcting for scattering by the disk and extinction by the envelope. The flux loss from scattering was calculated by utilizing \texttt{hyperion} RT models, and was found to be very large, and the flux loss from envelope extinction was calculated by optical depth estimates by fitting the rotation diagram of the H$_2$ 0-0 S(1-4) lines.
    \item We determined an accretion luminosity of $L_{acc} \sim 0.4~ \Lsun$. For L1527, $L_{tot} \sim 2.75 \Lsun$ which implies that most of the luminosity of the system is coming from the protostar, not accretion. 
    \item We calculate an accretion rate for L1527 of $\dot{M} \sim 1\times10^{-7}~\Lsun \text{yr}^{-1}$. This accretion rate is too low to explain the current measured mass for L1527, over a typical protostellar lifetime. indicating that the accretion rate must have been higher in the past, supporting theories of non-steady accretion.
    \item Detections of OH and water may be additional evidence for a UV field and therefore may provide further evidence for accretion. Shorter wavelength OH lines are needed to verify that the OH is in fact due to UV flux.
    \item The emission morphology of the Br$\alpha$ line and local continuum as well as the OH and water indicate a potential accretion asymmetry within the inner disk. These emission maps show that accretion may be stronger on the eastern side of the disk compared to the western side.

\end{itemize}

%% Acknowledgments section
\begin{acknowledgments}
We would like to thank the anonymous referee for their constructive review. This work is based on observations made with the NASA/ESA/CSA James Webb Space Telescope. The data were obtained from the Mikulski Archive for Space Telescopes at the Space Telescope Science Institute, which is operated by the Association of Universities for Research in Astronomy, Inc., under NASA contract NAS 5-03127 for JWST. These observations are associated with programs \#1798 and \#1290. Support for program \#1798 was provided by NASA through a grant from the Space Telescope Science Institute, which is operated by the Association of Universities for Research in Astronomy, Inc., under NASA contract NAS 5-03127. All the {\it JWST} data used in this paper can be found in MAST: \dataset[10.17909/3jvs-xp73]{http://dx.doi.org/10.17909/3jvs-xp73}.
P.D.S is supported by a National Science Foundation Astronomy \& Astrophysics Grant under Award No. 2305482.
Leiden is supported by funding from the European Research Council (ERC) under the European Union’s Horizon 2020 research and innovation program (grant agreement No. 291141 MOLDISK), and by NWO through TOP-1 grant 614.001.751.
R.D. acknowledges support from ERC advanced grant H2020-ER-2016-ADG-743029 EASY
\end{acknowledgments}


\vspace{5mm}
\facilities{JWST, ALMA}

%% Similar to \facility{}, there is the xoptional \software command to allow 

\software{\texttt{astropy} \citep{astropy2022}, \texttt{photutils} \citep{photutils}, \texttt{emcee} \citep{emcee}, \texttt{scipy} \citep{scipy2020}, \texttt{LMFIT} \citep{lmfit2016}, \texttt{hyperion} \citep{robitaille2011}, \texttt{stpsf} \citep{perrin2014}, \texttt{reproject} \citep{reproject}}



\FloatBarrier
\appendix
\section{Radiative Transfer Modeling}\label{appendix: RT}
\begin{figure}
    \centering
    \includegraphics[width=\linewidth]{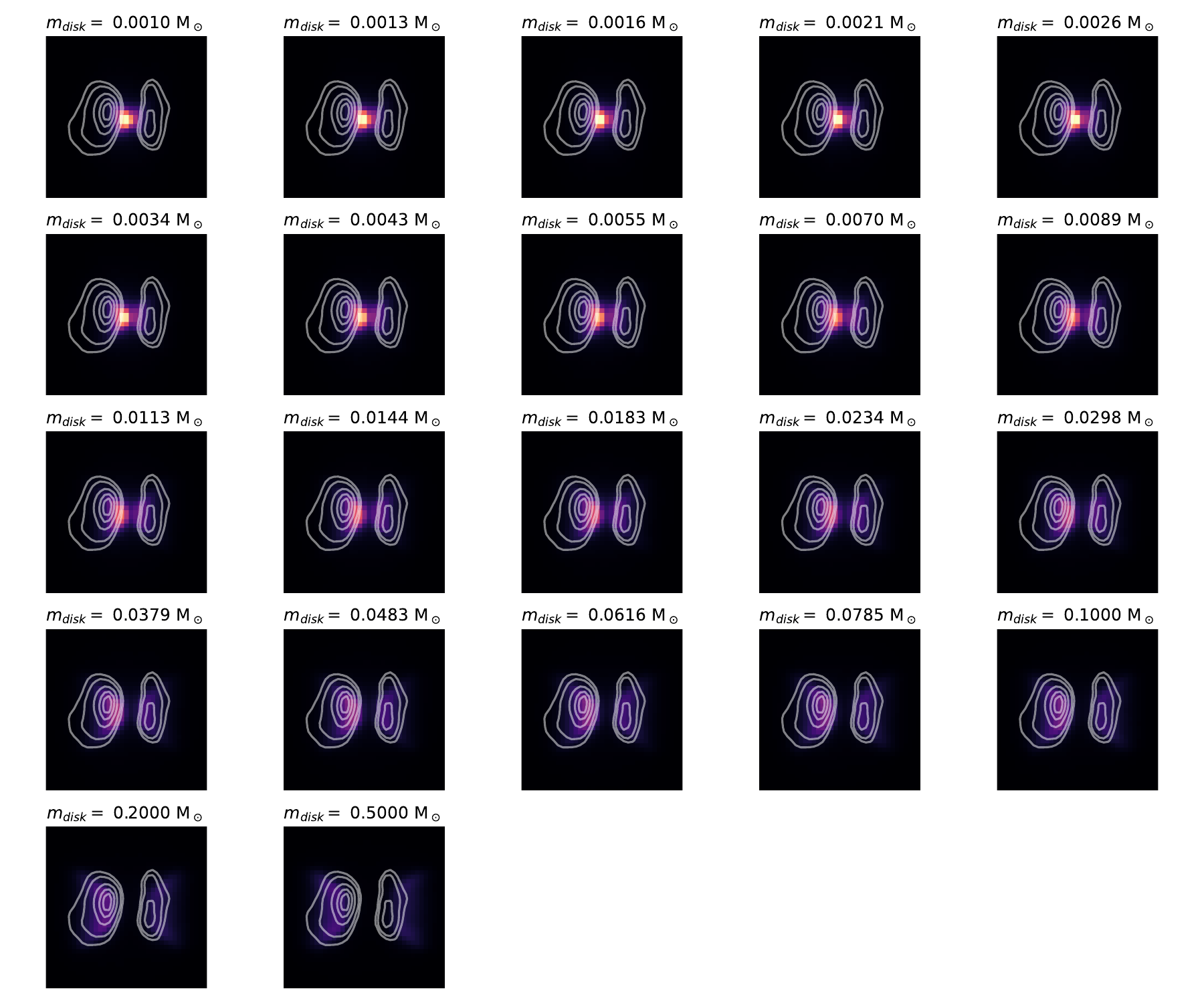}
    \caption{The disk + envelope RT models for all of the disk masses that we tested. The contours represent the NIRSpec 4 $\mu$m continuum and all of the images are shown on the same color-scale, which was set by the minimum and maximum flux of the NIRSpec 4 $\mu$m continuum.}
    \label{fig:rt grid}
\end{figure}

In Figure \ref{fig:rt model examples}, we show a few of the RT models generated with \texttt{hyperion} which we use to estimate the amount of intrinsic Br$\alpha$ emission observed in scattered light (before extinction from the envelope). The model parameters we used are shown in Table \ref{tab:rt parameters} and for completeness, we show all of the RT models in Figure \ref{fig:rt grid}.

\begin{table}[]
    \centering
    \begin{tabular}{l|l|c}
        \hline\hline
        Parameter & Description & Value(s) \\
        \hline
         R$_*$ (R$_\odot$) & Stellar radius & 2.09 \\
         T$_*$ (K) & Stellar temperature & 4000 \\
         L$_*$ (L$_\odot$) & System luminosity & 2.75 \\
         M$_*$ (M$_\odot$) & Stellar mass & 0.5 \\
         \hline
         M$_{disk}$ (M$_\odot$) & Total disk mass & 0.001 - 0.5 \\
         R$_{disk,min}$ (R$_*$) & Disk inner radius & 14.75 \\
         R$_{disk,max}$ (AU) & Disk outer radius & 125.0 \\
         H$_0$ (AU) & Scale height at 100 AU & 48.0 \\
         $p$ & Disk radial density exponent & -2.5 \\
         $\beta$ & Disk scale height exponent & 1.3 \\
         \hline
         R$_C$ (AU) & Envelope centrifugal radius & 125.0 \\
         R$_{env,min}$ (R$_*$) & Envelope inner radius & 42.75 \\
         R$_{env, max}$ (AU) & Envelope outer radius & 15000 \\
         $\dot{M}_{env}$ (M$_\odot$ yr$^{-1}$) & Envelope mass infall rate & $4.5\times 10^{-6}$ \\
         b$_{out}$ & Outer cavity shape exponent & 1.5 \\
         $\theta_{open,out}$ (\degree) & Outer cavity opening angle & 20 \\
         $\rho_c$ (g cm$^{-3}$) & Cavity density & 0 \\
         \hline
    \end{tabular}
    \caption{Summary of model parameters used in our RT modeling, based on the results from \citet{tobin2013}.}
    \label{tab:rt parameters}
\end{table}

\section{Extinction Estimation}\label{appendix: extinction}
\begin{figure}
    \centering
    \includegraphics[width=\linewidth]{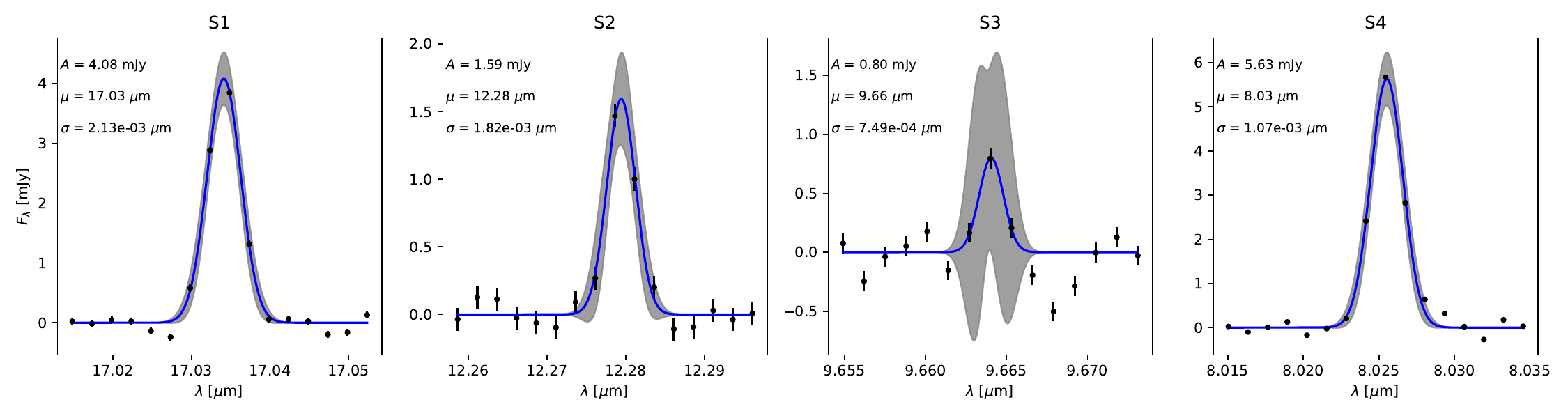}
    \caption{The line profiles and Gaussian fits of the H$_2$ S(1)-S(4) line used to construct the rotation diagram. The black points and error bars show the data and empirically estimated uncertainties, the blue line shows the best fit Gaussian model, and the gray region shows the 1$\sigma$ range on the model.}
    \label{fig:h2 fits}
\end{figure}

To create the rotation diagram shown in Figure \ref{fig:rot curve}, we first needed to measure the line fluxes of the H$_2$ 0-0 S(1-4) lines. We used the elliptical apertures shown in Figure \ref{fig:full_spec} to extract the line profiles since that is the spatial region where the Br$\alpha$ emission is concentrated. Thus, the H$_2$ within this aperture will probe the region of the envelope responsible for extincting the scattered light. In Figure \ref{fig:h2 fits}, we show the extracted H$_2$ line profiles and the Gaussian fits used to determine their associated line fluxes. 

\begin{figure}
    \centering
    \includegraphics[scale=0.7]{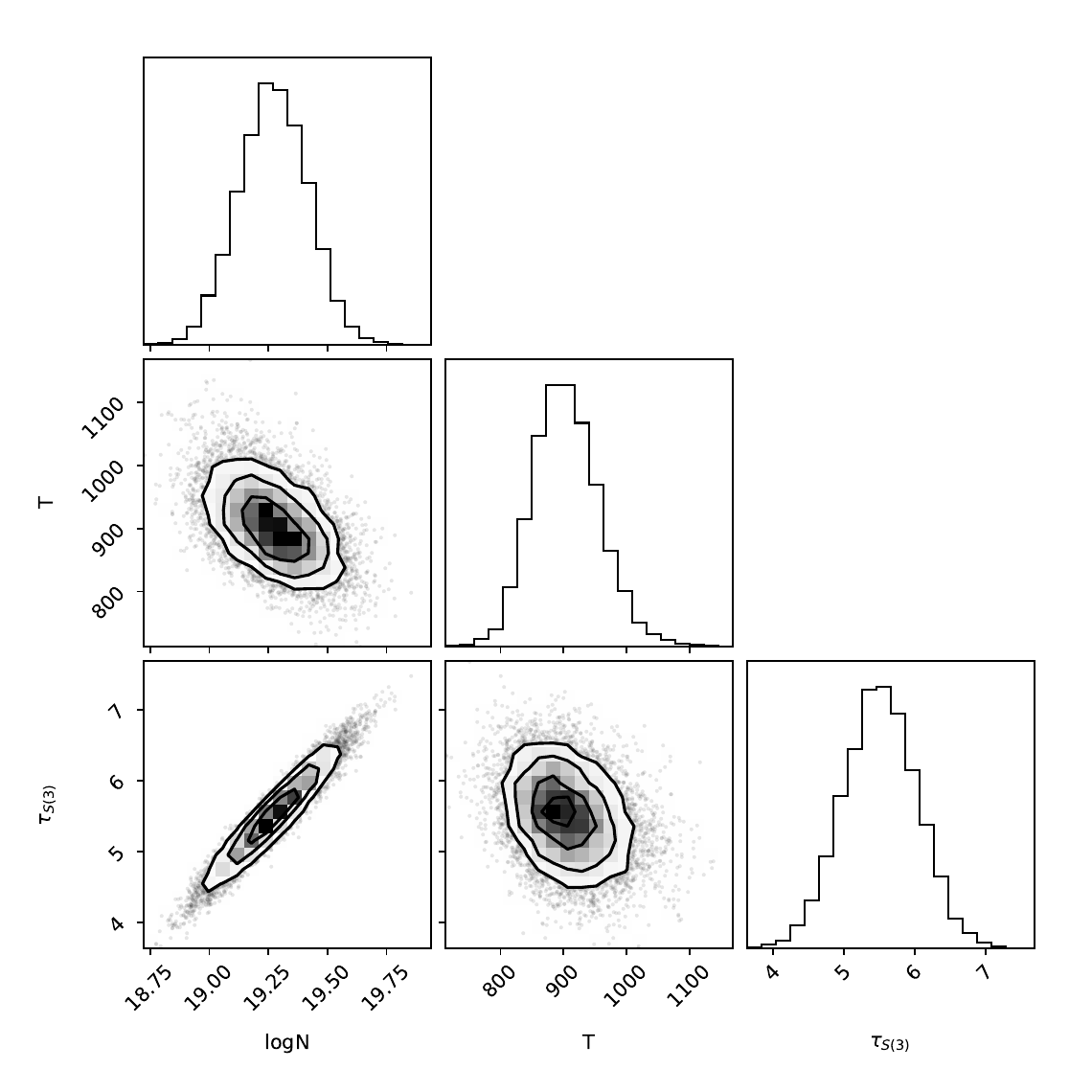}
    \caption{Corner plot showing the posterior distributions from the MCMC fit of the rotation diagram.}
    \label{fig:corner}
\end{figure}

We fit the resulting rotation diagram for the column density $N$, rotation temperature $T_{rot}$, and extinction at 9.7 $\mu$m, $\tau_{S3}$, via MCMC using the \texttt{emcee} package \citep{emcee}. We use this approach to better sample the posterior distribution on our parameters and constrain our uncertainties. We show this fit in Figure \ref{fig:rot curve}, and we show the corresponding corner plots in Figure \ref{fig:corner}.

\FloatBarrier
\bibliography{drechsler_bib}{}

\begin{thebibliography}{}
\expandafter\ifx\csname natexlab\endcsname\relax\def\natexlab#1{#1}\fi
\providecommand{\url}[1]{\href{#1}{#1}}
\providecommand{\dodoi}[1]{doi:~\href{http://doi.org/#1}{\nolinkurl{#1}}}
\providecommand{\doeprint}[1]{\href{http://ascl.net/#1}{\nolinkurl{http://ascl.net/#1}}}
\providecommand{\doarXiv}[1]{\href{https://arxiv.org/abs/#1}{\nolinkurl{https://arxiv.org/abs/#1}}}

% type= article
\bibitem[{P. {Andre} {et~al.}(1993){Andre}, {Ward-Thompson}, \&
  {Barsony}}]{andre1993}
{Andre}, P., {Ward-Thompson}, D., \& {Barsony}, M. 1993,
  \bibinfo{title}{{Submillimeter Continuum Observations of rho Ophiuchi A: The
  Candidate Protostar VLA 1623 and Prestellar Clumps},} \apj, 406, 122,
  \dodoi{10.1086/172425}

% type= inproceedings
\bibitem[{P. {Andre} {et~al.}(2000){Andre}, {Ward-Thompson}, \&
  {Barsony}}]{andre2000}
{Andre}, P., {Ward-Thompson}, D., \& {Barsony}, M. 2000, \bibinfo{title}{{From
  Prestellar Cores to Protostars: the Initial Conditions of Star Formation},}
  in Protostars and Planets IV, ed. V.~{Mannings}, A.~P. {Boss}, \& S.~S.
  {Russell}, 59, \dodoi{10.48550/arXiv.astro-ph/9903284}

% type= article
\bibitem[{ {Astropy Collaboration} {et~al.}(2022){Astropy Collaboration},
  {Price-Whelan}, {Lim}, {Earl}, {Starkman}, {Bradley}, {Shupe}, {Patil},
  {Corrales}, {Brasseur}, {N{\"o}the}, {Donath}, {Tollerud}, {Morris},
  {Ginsburg}, {Vaher}, {Weaver}, {Tocknell}, {Jamieson}, {van Kerkwijk},
  {Robitaille}, {Merry}, {Bachetti}, {G{\"u}nther}, {Aldcroft},
  {Alvarado-Montes}, {Archibald}, {B{\'o}di}, {Bapat}, {Barentsen},
  {Baz{\'a}n}, {Biswas}, {Boquien}, {Burke}, {Cara}, {Cara}, {Conroy},
  {Conseil}, {Craig}, {Cross}, {Cruz}, {D'Eugenio}, {Dencheva}, {Devillepoix},
  {Dietrich}, {Eigenbrot}, {Erben}, {Ferreira}, {Foreman-Mackey}, {Fox},
  {Freij}, {Garg}, {Geda}, {Glattly}, {Gondhalekar}, {Gordon}, {Grant},
  {Greenfield}, {Groener}, {Guest}, {Gurovich}, {Handberg}, {Hart},
  {Hatfield-Dodds}, {Homeier}, {Hosseinzadeh}, {Jenness}, {Jones}, {Joseph},
  {Kalmbach}, {Karamehmetoglu}, {Ka{\l}uszy{\'n}ski}, {Kelley}, {Kern},
  {Kerzendorf}, {Koch}, {Kulumani}, {Lee}, {Ly}, {Ma}, {MacBride}, {Maljaars},
  {Muna}, {Murphy}, {Norman}, {O'Steen}, {Oman}, {Pacifici}, {Pascual},
  {Pascual-Granado}, {Patil}, {Perren}, {Pickering}, {Rastogi}, {Roulston},
  {Ryan}, {Rykoff}, {Sabater}, {Sakurikar}, {Salgado}, {Sanghi}, {Saunders},
  {Savchenko}, {Schwardt}, {Seifert-Eckert}, {Shih}, {Jain}, {Shukla}, {Sick},
  {Simpson}, {Singanamalla}, {Singer}, {Singhal}, {Sinha}, {Sip{\H{o}}cz},
  {Spitler}, {Stansby}, {Streicher}, {{\v{S}}umak}, {Swinbank}, {Taranu},
  {Tewary}, {Tremblay}, {de Val-Borro}, {Van Kooten}, {Vasovi{\'c}}, {Verma},
  {de Miranda Cardoso}, {Williams}, {Wilson}, {Winkel}, {Wood-Vasey}, {Xue},
  {Yoachim}, {Zhang}, {Zonca}, \& {Astropy Project Contributors}}]{astropy2022}
{Astropy Collaboration}, {Price-Whelan}, A.~M., {Lim}, P.~L., {et~al.} 2022,
  \bibinfo{title}{{The Astropy Project: Sustaining and Growing a
  Community-oriented Open-source Project and the Latest Major Release (v5.0) of
  the Core Package},} \apj, 935, 167, \dodoi{10.3847/1538-4357/ac7c74}

% type= article
\bibitem[{T. {B{\"o}ker} {et~al.}(2022){B{\"o}ker}, {Arribas},
  {L{\"u}tzgendorf}, {Alves de Oliveira}, {Beck}, {Birkmann}, {Bunker},
  {Charlot}, {de Marchi}, {Ferruit}, {Giardino}, {Jakobsen}, {Kumari},
  {L{\'o}pez-Caniego}, {Maiolino}, {Manjavacas}, {Marston}, {Moseley},
  {Muzerolle}, {Ogle}, {Pirzkal}, {Rauscher}, {Rawle}, {Rix}, {Sabbi},
  {Sargent}, {Sirianni}, {te Plate}, {Valenti}, {Willott}, \&
  {Zeidler}}]{boker2022}
{B{\"o}ker}, T., {Arribas}, S., {L{\"u}tzgendorf}, N., {et~al.} 2022,
  \bibinfo{title}{{The Near-Infrared Spectrograph (NIRSpec) on the James Webb
  Space Telescope. III. Integral-field spectroscopy},} \aap, 661, A82,
  \dodoi{10.1051/0004-6361/202142589}

% type= article
\bibitem[{S. {Bontemps} {et~al.}(1996){Bontemps}, {Andre}, {Terebey}, \&
  {Cabrit}}]{bontemps1996}
{Bontemps}, S., {Andre}, P., {Terebey}, S., \& {Cabrit}, S. 1996,
  \bibinfo{title}{{Evolution of outflow activity around low-mass embedded young
  stellar objects},} \aap, 311, 858

% type= inproceedings
\bibitem[{J. {Bouvier} {et~al.}(2007){Bouvier}, {Alencar}, {Harries},
  {Johns-Krull}, \& {Romanova}}]{bouvier2007}
{Bouvier}, J., {Alencar}, S.~H.~P., {Harries}, T.~J., {Johns-Krull}, C.~M., \&
  {Romanova}, M.~M. 2007, \bibinfo{title}{{Magnetospheric Accretion in
  Classical T Tauri Stars},} in Protostars and Planets V, ed. B.~{Reipurth},
  D.~{Jewitt}, \& K.~{Keil}, 479, \dodoi{10.48550/arXiv.astro-ph/0603498}

% type= misc
\bibitem[{L. Bradley {et~al.}(2025)Bradley, Sip{\H o}cz, Robitaille, Tollerud,
  Vin{\'{\i}}cius, Deil, Barbary, Wilson, Busko, Donath, G{\"u}nther, Cara,
  Lim, Me{\ss}linger, Burnett, Conseil, Droettboom, Bostroem, Bray, Bratholm,
  Jamieson, Ginsburg, Barentsen, Craig, Pascual, Rathi, Perrin, \&
  Morris}]{photutils}
Bradley, L., Sip{\H o}cz, B., Robitaille, T., {et~al.} 2025, astropy/photutils:
  2.2.0, 2.2.0 Zenodo, \dodoi{10.5281/zenodo.14889440}

% type= article
\bibitem[{N. {Calvet} \& E. {Gullbring}(1998){Calvet} \&
  {Gullbring}}]{calvet_gullbring1998}
{Calvet}, N., \& {Gullbring}, E. 1998, \bibinfo{title}{{The Structure and
  Emission of the Accretion Shock in T Tauri Stars},} \apj, 509, 802,
  \dodoi{10.1086/306527}

% type= article
\bibitem[{P. {Cassen} \& A. {Moosman}(1981){Cassen} \&
  {Moosman}}]{cassen&moosman1981}
{Cassen}, P., \& {Moosman}, A. 1981, \bibinfo{title}{{On the formation of
  protostellar disks},} \icarus, 48, 353, \dodoi{10.1016/0019-1035(81)90051-8}

% type= article
\bibitem[{C.~J. {Chandler} \& J.~S. {Richer}(2000){Chandler} \&
  {Richer}}]{chandler&richer2000}
{Chandler}, C.~J., \& {Richer}, J.~S. 2000, \bibinfo{title}{{The Structure of
  Protostellar Envelopes Derived from Submillimeter Continuum Images},} \apj,
  530, 851, \dodoi{10.1086/308401}

% type= article
\bibitem[{J. {Cleaver} {et~al.}(2023){Cleaver}, {Hartmann}, \&
  {Bae}}]{cleaver2023}
{Cleaver}, J., {Hartmann}, L., \& {Bae}, J. 2023, \bibinfo{title}{{Magnetically
  activated accretion outbursts of pre-main-sequence discs},} \mnras, 523,
  5522, \dodoi{10.1093/mnras/stad1784}

% type= article
\bibitem[{M.~S. {Connelley} \& T.~P. {Greene}(2014){Connelley} \&
  {Greene}}]{connelley2014}
{Connelley}, M.~S., \& {Greene}, T.~P. 2014, \bibinfo{title}{{Near-IR
  Spectroscopic Monitoring of Class I Protostars: Variability of Accretion and
  Wind Indicators},} \aj, 147, 125, \dodoi{10.1088/0004-6256/147/6/125}

% type= article
\bibitem[{V. {Delabrosse} {et~al.}(2024){Delabrosse}, {Dougados}, {Cabrit},
  {Tabone}, {Tychoniec}, {Ray}, {Podio}, \& {McClure}}]{delabrosse2024}
{Delabrosse}, V., {Dougados}, C., {Cabrit}, S., {et~al.} 2024,
  \bibinfo{title}{{JWST study of the DG Tau B disk-wind candidate. I. Overview
  and nested H$_{2}$-CO outflows},} \aap, 688, A173,
  \dodoi{10.1051/0004-6361/202449176}

% type= article
\bibitem[{M.~M. {Dunham} {et~al.}(2010){Dunham}, {Evans}, {Terebey},
  {Dullemond}, \& {Young}}]{dunham2010}
{Dunham}, M.~M., {Evans}, II, N.~J., {Terebey}, S., {Dullemond}, C.~P., \&
  {Young}, C.~H. 2010, \bibinfo{title}{{Evolutionary Signatures in the
  Formation of Low-Mass Protostars. II. Toward Reconciling Models and
  Observations},} \apj, 710, 470, \dodoi{10.1088/0004-637X/710/1/470}

% type= article
\bibitem[{S.~A. {Federman} {et~al.}(2024){Federman}, {Megeath}, {Rubinstein},
  {Gutermuth}, {Narang}, {Tyagi}, {Manoj}, {Anglada}, {Atnagulov}, {Beuther},
  {Bourke}, {Brunken}, {Caratti o Garatti}, {Evans}, {Fischer}, {Furlan},
  {Green}, {Habel}, {Hartmann}, {Karnath}, {Klaassen}, {Linz}, {Looney},
  {Osorio}, {Muzerolle Page}, {Nazari}, {Pokhrel}, {Rahatgaonkar}, {Rocha},
  {Sheehan}, {Slavicinska}, {Stanke}, {Stutz}, {Tobin}, {Tychoniec}, {Van
  Dishoeck}, {Watson}, {Wolk}, \& {Yang}}]{Federman2024}
{Federman}, S.~A., {Megeath}, S.~T., {Rubinstein}, A.~E., {et~al.} 2024,
  \bibinfo{title}{{Investigating Protostellar Accretion-driven Outflows across
  the Mass Spectrum: JWST NIRSpec Integral Field Unit 3{\textendash}5
  {\ensuremath{\mu}}m Spectral Mapping of Five Young Protostars},} \apj, 966,
  41, \dodoi{10.3847/1538-4357/ad2fa0}

% type= article
\bibitem[{E. {Fiorellino} {et~al.}(2023){Fiorellino}, {Tychoniec},
  {Cruz-S{\'a}enz de Miera}, {Antoniucci}, {K{\'o}sp{\'a}l}, {Manara},
  {Nisini}, \& {Rosotti}}]{fiorellino2023}
{Fiorellino}, E., {Tychoniec}, {\L}., {Cruz-S{\'a}enz de Miera}, F., {et~al.}
  2023, \bibinfo{title}{{The Mass Accretion Rate and Stellar Properties in
  Class I Protostars},} \apj, 944, 135, \dodoi{10.3847/1538-4357/aca320}

% type= inproceedings
\bibitem[{W.~J. {Fischer} {et~al.}(2023){Fischer}, {Hillenbrand}, {Herczeg},
  {Johnstone}, {Kospal}, \& {Dunham}}]{fischer2023}
{Fischer}, W.~J., {Hillenbrand}, L.~A., {Herczeg}, G.~J., {et~al.} 2023,
  \bibinfo{title}{{Accretion Variability as a Guide to Stellar Mass Assembly},}
  in Astronomical Society of the Pacific Conference Series, Vol. 534,
  Protostars and Planets VII, ed. S.~{Inutsuka}, Y.~{Aikawa}, T.~{Muto},
  K.~{Tomida}, \& M.~{Tamura}, 355, \dodoi{10.48550/arXiv.2203.11257}

% type= article
\bibitem[{D. {Foreman-Mackey} {et~al.}(2013){Foreman-Mackey}, {Hogg}, {Lang},
  \& {Goodman}}]{emcee}
{Foreman-Mackey}, D., {Hogg}, D.~W., {Lang}, D., \& {Goodman}, J. 2013,
  \bibinfo{title}{{emcee: The MCMC Hammer},} \pasp, 125, 306,
  \dodoi{10.1086/670067}

% type= article
\bibitem[{L. {Francis} {et~al.}(2025){Francis}, {van Dishoeck}, {Caratti o
  Garatti}, {van Gelder}, {Gieser}, {Beuther}, {Ray}, {Tychoniec}, {Nazari},
  {Reyes}, {Kavanagh}, {Klaassen}, {G{\"u}del}, \& {Henning}}]{francis2025}
{Francis}, L., {van Dishoeck}, E.~F., {Caratti o Garatti}, A., {et~al.} 2025,
  \bibinfo{title}{{JOYS: The [D/H] abundance derived from protostellar outflows
  across the Galactic disk measured with JWST},} \aap, 694, A174,
  \dodoi{10.1051/0004-6361/202451629}

% type= article
\bibitem[{B.~A.~L. {Gaches} {et~al.}(2024){Gaches}, {Tan}, {Rosen}, \&
  {Kuiper}}]{gaches2024}
{Gaches}, B. A.~L., {Tan}, J.~C., {Rosen}, A.~L., \& {Kuiper}, R. 2024,
  \bibinfo{title}{{The High-resolution Accretion Disks of Embedded protoStars
  (HADES) simulations: I. Impact of protostellar magnetic fields on accretion
  modes},} \aap, 692, A219, \dodoi{10.1051/0004-6361/202451842}

% type= article
\bibitem[{D. {Harsono} {et~al.}(2023){Harsono}, {Bjerkeli}, {Ramsey},
  {Pontoppidan}, {Kristensen}, {J{\o}rgensen}, {Calcutt}, {Li}, \&
  {Plunkett}}]{Harsono2023}
{Harsono}, D., {Bjerkeli}, P., {Ramsey}, J.~P., {et~al.} 2023,
  \bibinfo{title}{{JWST Peers into the Class I Protostar TMC1A: Atomic Jet and
  Spatially Resolved Dissociative Shock Region},} \apjl, 951, L32,
  \dodoi{10.3847/2041-8213/acdfca}

% type= article
\bibitem[{L. {Hartmann} {et~al.}(1994){Hartmann}, {Hewett}, \&
  {Calvet}}]{hartmann1994}
{Hartmann}, L., {Hewett}, R., \& {Calvet}, N. 1994,
  \bibinfo{title}{{Magnetospheric Accretion Models for T Tauri Stars. I. Balmer
  Line Profiles without Rotation},} \apj, 426, 669, \dodoi{10.1086/174104}

% type= article
\bibitem[{L. {Hartmann} {et~al.}(2025){Hartmann}, {Tobin}, {Sheehan},
  {Kounkel}, \& {Zhao}}]{hartmann2025}
{Hartmann}, L., {Tobin}, J.~J., {Sheehan}, P., {Kounkel}, M., \& {Zhao}, C.
  2025, \bibinfo{title}{{On the protostellar mass-luminosity relation},} arXiv
  e-prints, arXiv:2507.18728.
\newblock \doarXiv{2507.18728}

% type= article
\bibitem[{M.~R. {Hogerheijde} {et~al.}(1998){Hogerheijde}, {van Dishoeck},
  {Blake}, \& {van Langevelde}}]{hogerheijde1998}
{Hogerheijde}, M.~R., {van Dishoeck}, E.~F., {Blake}, G.~A., \& {van
  Langevelde}, H.~J. 1998, \bibinfo{title}{{Envelope Structure on 700 AU Scales
  and the Molecular Outflows of Low-Mass Young Stellar Objects},} \apj, 502,
  315, \dodoi{10.1086/305885}

% type= article
\bibitem[{P. {Jakobsen} {et~al.}(2022){Jakobsen}, {Ferruit}, {Alves de
  Oliveira}, {Arribas}, {Bagnasco}, {Barho}, {Beck}, {Birkmann}, {B{\"o}ker},
  {Bunker}, {Charlot}, {de Jong}, {de Marchi}, {Ehrenwinkler}, {Falcolini},
  {Fels}, {Franx}, {Franz}, {Funke}, {Giardino}, {Gnata}, {Holota}, {Honnen},
  {Jensen}, {Jentsch}, {Johnson}, {Jollet}, {Karl}, {Kling}, {K{\"o}hler},
  {Kolm}, {Kumari}, {Lander}, {Lemke}, {L{\'o}pez-Caniego}, {L{\"u}tzgendorf},
  {Maiolino}, {Manjavacas}, {Marston}, {Maschmann}, {Maurer}, {Messerschmidt},
  {Moseley}, {Mosner}, {Mott}, {Muzerolle}, {Pirzkal}, {Pittet}, {Plitzke},
  {Posselt}, {Rapp}, {Rauscher}, {Rawle}, {Rix}, {R{\"o}del}, {Rumler},
  {Sabbi}, {Salvignol}, {Schmid}, {Sirianni}, {Smith}, {Strada}, {te Plate},
  {Valenti}, {Wettemann}, {Wiehe}, {Wiesmayer}, {Willott}, {Wright}, {Zeidler},
  \& {Zincke}}]{jakobsen2022}
{Jakobsen}, P., {Ferruit}, P., {Alves de Oliveira}, C., {et~al.} 2022,
  \bibinfo{title}{{The Near-Infrared Spectrograph (NIRSpec) on the James Webb
  Space Telescope. I. Overview of the instrument and its capabilities},} \aap,
  661, A80, \dodoi{10.1051/0004-6361/202142663}

% type= article
\bibitem[{S.~J. {Kenyon} {et~al.}(1990){Kenyon}, {Hartmann}, {Strom}, \&
  {Strom}}]{kenyon1990}
{Kenyon}, S.~J., {Hartmann}, L.~W., {Strom}, K.~M., \& {Strom}, S.~E. 1990,
  \bibinfo{title}{{An IRAS Survey of the Taurus-Auriga Molecular Cloud},} \aj,
  99, 869, \dodoi{10.1086/115380}

% type= article
\bibitem[{O. {Komarova} \& W.~J. {Fischer}(2020){Komarova} \&
  {Fischer}}]{Komarova&Fischer2020}
{Komarova}, O., \& {Fischer}, W.~J. 2020, \bibinfo{title}{{Calibration of
  Brackett Alpha as an Accretion Indicator in T Tauri Stars},} Research Notes
  of the American Astronomical Society, 4, 6, \dodoi{10.3847/2515-5172/ab67bb}

% type= article
\bibitem[{L.~E. {Kristensen} {et~al.}(2012){Kristensen}, {van Dishoeck},
  {Bergin}, {Visser}, {Y{\i}ld{\i}z}, {San Jose-Garcia}, {J{\o}rgensen},
  {Herczeg}, {Johnstone}, {Wampfler}, \& et~al.}]{kristensen2012}
{Kristensen}, L.~E., {van Dishoeck}, E.~F., {Bergin}, E.~A., {et~al.} 2012,
  \bibinfo{title}{{Water in star-forming regions with Herschel (WISH). II.
  Evolution of 557 GHz {}1$_{10}$-{}1$_{01}$ emission in low-mass protostars},}
  \aap, 542, A8, \dodoi{10.1051/0004-6361/201118146}

% type= article
\bibitem[{E. {Laos} {et~al.}(2021){Laos}, {Greene}, {Najita}, \&
  {Stassun}}]{laos2021}
{Laos}, E., {Greene}, T.~P., {Najita}, J.~R., \& {Stassun}, K.~G. 2021,
  \bibinfo{title}{{The Near-stellar Environment of Class 0 Protostars: A First
  Look with Near-infrared Spectroscopy},} \apj, 921, 110,
  \dodoi{10.3847/1538-4357/ac1f1b}

% type= article
\bibitem[{V.~J.~M. {Le Gouellec} {et~al.}(2024){Le Gouellec}, {Greene},
  {Hillenbrand}, \& {Yates}}]{leGouellec2024}
{Le Gouellec}, V. J.~M., {Greene}, T.~P., {Hillenbrand}, L.~A., \& {Yates}, Z.
  2024, \bibinfo{title}{{New Insights on the Accretion Properties of Class 0
  Protostars from 2 {\ensuremath{\mu}}m Spectroscopy},} \apj, 966, 91,
  \dodoi{10.3847/1538-4357/ad2935}

% type= article
\bibitem[{V.~J.~M. {Le Gouellec} {et~al.}(2025){Le Gouellec}, {Lew}, {Greene},
  {Johnstone}, {Gusdorf}, {Francis}, {DeWitt}, {Meyer}, {Tychoniec}, {van
  Dishoeck}, \& et~al.}]{leGouellec2025}
{Le Gouellec}, V. J.~M., {Lew}, B. W.~P., {Greene}, T.~P., {et~al.} 2025,
  \bibinfo{title}{{Unveiling Two Deeply Embedded Young Protostars in the S68N
  Class 0 Protostellar Core with JWST/NIRSpec},} \apj, 985, 225,
  \dodoi{10.3847/1538-4357/adcac4}

% type= article
\bibitem[{K.~L. {Luhman}(2018){Luhman}}]{luhman2018}
{Luhman}, K.~L. 2018, \bibinfo{title}{{The Stellar Membership of the Taurus
  Star-forming Region},} \aj, 156, 271, \dodoi{10.3847/1538-3881/aae831}

% type= article
\bibitem[{M.~K. {McClure} {et~al.}(2013){McClure}, {Calvet}, {Espaillat},
  {Hartmann}, {Hern{\'a}ndez}, {Ingleby}, {Luhman}, {D'Alessio}, \&
  {Sargent}}]{mcclure2013}
{McClure}, M.~K., {Calvet}, N., {Espaillat}, C., {et~al.} 2013,
  \bibinfo{title}{{Characterizing the Stellar Photospheres and Near-infrared
  Excesses in Accreting T Tauri Systems},} \apj, 769, 73,
  \dodoi{10.1088/0004-637X/769/1/73}

% type= article
\bibitem[{M. {Narang} {et~al.}(2024){Narang}, {Manoj}, {Tyagi}, {Watson},
  {Megeath}, {Federman}, {Rubinstein}, {Gutermuth}, {Caratti o Garatti},
  {Beuther}, {Bourke}, {Van Dishoeck}, {Evans}, {Anglada}, {Osorio}, {Stanke},
  {Muzerolle}, {Looney}, {Yang}, {Klaassen}, {Karnath}, {Atnagulov}, {Brunken},
  {Fischer}, {Furlan}, {Green}, {Habel}, {Hartmann}, {Linz}, {Nazari},
  {Pokhrel}, {Rahatgaonkar}, {Rocha}, {Sheehan}, {Slavicinska}, {Stutz},
  {Tobin}, {Tychoniec}, \& {Wolk}}]{narang2024}
{Narang}, M., {Manoj}, P., {Tyagi}, H., {et~al.} 2024,
  \bibinfo{title}{{Discovery of a Collimated Jet from the Low-luminosity
  Protostar IRAS 16253‑2429 in a Quiescent Accretion Phase with the JWST},}
  \apjl, 962, L16, \dodoi{10.3847/2041-8213/ad1de3}

% type= article
\bibitem[{A. {Natta} {et~al.}(2001){Natta}, {Prusti}, {Neri}, {Wooden},
  {Grinin}, \& {Mannings}}]{natta2001}
{Natta}, A., {Prusti}, T., {Neri}, R., {et~al.} 2001, \bibinfo{title}{{A
  reconsideration of disk properties in Herbig Ae stars},} \aap, 371, 186,
  \dodoi{10.1051/0004-6361:20010334}

% type= misc
\bibitem[{M. {Newville} {et~al.}(2016){Newville}, {Stensitzki}, {Allen},
  {Rawlik}, {Ingargiola}, \& {Nelson}}]{lmfit2016}
{Newville}, M., {Stensitzki}, T., {Allen}, D.~B., {et~al.} 2016, {Lmfit:
  Non-Linear Least-Square Minimization and Curve-Fitting for Python},,
  Astrophysics Source Code Library, record ascl:1606.014 \doeprint{1606.014}

% type= inproceedings
\bibitem[{S.~S.~R. {Offner} {et~al.}(2023){Offner}, {Moe}, {Kratter},
  {Sadavoy}, {Jensen}, \& {Tobin}}]{offner2023}
{Offner}, S.~S.~R., {Moe}, M., {Kratter}, K.~M., {et~al.} 2023,
  \bibinfo{title}{{The Origin and Evolution of Multiple Star Systems},} in
  Astronomical Society of the Pacific Conference Series, Vol. 534, Protostars
  and Planets VII, ed. S.~{Inutsuka}, Y.~{Aikawa}, T.~{Muto}, K.~{Tomida}, \&
  M.~{Tamura}, 275, \dodoi{10.48550/arXiv.2203.10066}

% type= article
\bibitem[{N. {Ohashi} {et~al.}(2014){Ohashi}, {Saigo}, {Aso}, {Aikawa},
  {Koyamatsu}, {Machida}, {Saito}, {Takahashi}, {Takakuwa}, {Tomida},
  {Tomisaka}, \& {Yen}}]{ohashi2014}
{Ohashi}, N., {Saigo}, K., {Aso}, Y., {et~al.} 2014, \bibinfo{title}{{Formation
  of a Keplerian Disk in the Infalling Envelope around L1527 IRS:
  Transformation from Infalling Motions to Kepler Motions},} \apj, 796, 131,
  \dodoi{10.1088/0004-637X/796/2/131}

% type= article
\bibitem[{N. {Ohashi} {et~al.}(2023){Ohashi}, {Tobin}, {J{\o}rgensen},
  {Takakuwa}, {Sheehan}, {Aikawa}, {Li}, {Looney}, {Williams}, {Aso}, {Sharma},
  {Sai}, {Yamato}, {Lee}, {Tomida}, {Yen}, {Encalada}, {Flores}, {Gavino},
  {Kido}, {Han}, {Lin}, {Narayanan}, {Phuong}, {Santamar{\'\i}a-Miranda},
  {Thieme}, {van't Hoff}, {de Gregorio-Monsalvo}, {Koch}, {Kwon}, {Lai}, {Lee},
  {Plunkett}, {Saigo}, {Hirano}, {Lam}, \& {Mori}}]{ohashi2023}
{Ohashi}, N., {Tobin}, J.~J., {J{\o}rgensen}, J.~K., {et~al.} 2023,
  \bibinfo{title}{{Early Planet Formation in Embedded Disks (eDisk). I.
  Overview of the Program and First Results},} \apj, 951, 8,
  \dodoi{10.3847/1538-4357/acd384}

% type= inproceedings
\bibitem[{M.~D. {Perrin} {et~al.}(2014){Perrin}, {Sivaramakrishnan}, {Lajoie},
  {Elliott}, {Pueyo}, {Ravindranath}, \& {Albert}}]{perrin2014}
{Perrin}, M.~D., {Sivaramakrishnan}, A., {Lajoie}, C.-P., {et~al.} 2014,
  \bibinfo{title}{{Updated point spread function simulations for JWST with
  WebbPSF},} in Society of Photo-Optical Instrumentation Engineers (SPIE)
  Conference Series, Vol. 9143, Space Telescopes and Instrumentation 2014:
  Optical, Infrared, and Millimeter Wave, ed. J.~M. {Oschmann}, Jr.,
  M.~{Clampin}, G.~G. {Fazio}, \& H.~A. {MacEwen}, 91433X,
  \dodoi{10.1117/12.2056689}

% type= article
\bibitem[{K.~M. {Pontoppidan} {et~al.}(2024){Pontoppidan}, {Evans}, {Bergner},
  \& {Yang}}]{pontoppidan2024}
{Pontoppidan}, K.~M., {Evans}, N., {Bergner}, J., \& {Yang}, Y.-L. 2024,
  \bibinfo{title}{{A Constrained Dust Opacity for Models of Dense Clouds and
  Protostellar Envelopes},} Research Notes of the American Astronomical
  Society, 8, 68, \dodoi{10.3847/2515-5172/ad303f}

% type= article
\bibitem[{G.~H. {Rieke} {et~al.}(2015){Rieke}, {Wright}, {B{\"o}ker},
  {Bouwman}, {Colina}, {Glasse}, {Gordon}, {Greene}, {G{\"u}del}, {Henning},
  {Justtanont}, {Lagage}, {Meixner}, {N{\o}rgaard-Nielsen}, {Ray}, {Ressler},
  {van Dishoeck}, \& {Waelkens}}]{rieke2015}
{Rieke}, G.~H., {Wright}, G.~S., {B{\"o}ker}, T., {et~al.} 2015,
  \bibinfo{title}{{The Mid-Infrared Instrument for the James Webb Space
  Telescope, I: Introduction},} \pasp, 127, 584, \dodoi{10.1086/682252}

% type= misc
\bibitem[{T. {Robitaille}(2018){Robitaille}}]{reproject}
{Robitaille}, T. 2018, {reproject: astronomical image reprojection in Python},
  v0.4 Zenodo, \dodoi{10.5281/zenodo.1162674}

% type= article
\bibitem[{T.~P. {Robitaille}(2011){Robitaille}}]{robitaille2011}
{Robitaille}, T.~P. 2011, \bibinfo{title}{{HYPERION: an open-source
  parallelized three-dimensional dust continuum radiative transfer code},}
  \aap, 536, A79, \dodoi{10.1051/0004-6361/201117150}

% type= article
\bibitem[{A.~C. {Rodriguez} \& L.~A. {Hillenbrand}(2022){Rodriguez} \&
  {Hillenbrand}}]{rodriguez&hillenbrand2022}
{Rodriguez}, A.~C., \& {Hillenbrand}, L.~A. 2022, \bibinfo{title}{{Application
  of a Steady-state Accretion Disk Model to Spectrophotometry and
  High-resolution Spectra of Two Recent FU Ori Outbursts},} \apj, 927, 144,
  \dodoi{10.3847/1538-4357/ac496b}

% type= article
\bibitem[{K. {Slavicinska} {et~al.}(2025){Slavicinska}, {Tychoniec}, {Navarro},
  {van Dishoeck}, {Tobin}, {van Gelder}, {Chen}, {Boogert}, {Drechsler},
  {Beuther}, {Caratti o Garatti}, {Megeath}, {Klaassen}, {Looney}, {Kavanagh},
  {Brunken}, {Sheehan}, \& {Fischer}}]{slavicinska2025}
{Slavicinska}, K., {Tychoniec}, {\L}., {Navarro}, M.~G., {et~al.} 2025,
  \bibinfo{title}{{HDO Ice Detected toward an Isolated Low-mass Protostar with
  JWST},} \apjl, 986, L19, \dodoi{10.3847/2041-8213/addb45}

% type= article
\bibitem[{B. {Tabone} {et~al.}(2021){Tabone}, {van Hemert}, {van Dishoeck}, \&
  {Black}}]{tabone2021}
{Tabone}, B., {van Hemert}, M.~C., {van Dishoeck}, E.~F., \& {Black}, J.~H.
  2021, \bibinfo{title}{{OH mid-infrared emission as a diagnostic of H$_{2}$O
  UV photodissociation. I. Model and application to the HH 211 shock},} \aap,
  650, A192, \dodoi{10.1051/0004-6361/202039549}

% type= article
\bibitem[{S. {Terebey} {et~al.}(1984){Terebey}, {Shu}, \&
  {Cassen}}]{terebey1984}
{Terebey}, S., {Shu}, F.~H., \& {Cassen}, P. 1984, \bibinfo{title}{{The
  collapse of the cores of slowly rotating isothermal clouds},} \apj, 286, 529,
  \dodoi{10.1086/162628}

% type= article
\bibitem[{J.~J. {Tobin} {et~al.}(2008){Tobin}, {Hartmann}, {Calvet}, \&
  {D'Alessio}}]{tobin2008}
{Tobin}, J.~J., {Hartmann}, L., {Calvet}, N., \& {D'Alessio}, P. 2008,
  \bibinfo{title}{{Constraining the Envelope Structure of L1527 IRS: Infrared
  Scattered Light Modeling},} \apj, 679, 1364, \dodoi{10.1086/587683}

% type= article
\bibitem[{J.~J. {Tobin} {et~al.}(2012){Tobin}, {Hartmann}, {Chiang}, {Wilner},
  {Looney}, {Loinard}, {Calvet}, \& {D'Alessio}}]{tobin2012}
{Tobin}, J.~J., {Hartmann}, L., {Chiang}, H.-F., {et~al.} 2012,
  \bibinfo{title}{{A \raisebox{-0.5ex}\textasciitilde0.2-solar-mass protostar
  with a Keplerian disk in the very young L1527 IRS system},} \nat, 492, 83,
  \dodoi{10.1038/nature11610}

% type= article
\bibitem[{J.~J. {Tobin} {et~al.}(2013){Tobin}, {Hartmann}, {Chiang}, {Wilner},
  {Looney}, {Loinard}, {Calvet}, \& {D'Alessio}}]{tobin2013}
{Tobin}, J.~J., {Hartmann}, L., {Chiang}, H.-F., {et~al.} 2013,
  \bibinfo{title}{{Modeling the Resolved Disk around the Class 0 Protostar
  L1527},} \apj, 771, 48, \dodoi{10.1088/0004-637X/771/1/48}

% type= article
\bibitem[{J.~J. {Tobin} {et~al.}(2010){Tobin}, {Hartmann}, \&
  {Loinard}}]{tobin2010}
{Tobin}, J.~J., {Hartmann}, L., \& {Loinard}, L. 2010, \bibinfo{title}{{The
  Inner Envelope and Disk of L1527 Revealed: Gemini L'-band-scattered Light
  Imaging},} \apjl, 722, L12, \dodoi{10.1088/2041-8205/722/1/L12}

% type= article
\bibitem[{J.~J. {Tobin} \& P.~D. {Sheehan}(2024){Tobin} \&
  {Sheehan}}]{tobin&sheehan2024}
{Tobin}, J.~J., \& {Sheehan}, P.~D. 2024, \bibinfo{title}{{An Observational
  View of Structure in Protostellar Systems},} \araa, 62, 203,
  \dodoi{10.1146/annurev-astro-052920-103752}

% type= article
\bibitem[{J.~J. {Tobin} {et~al.}(2011){Tobin}, {Hartmann}, {Chiang}, {Looney},
  {Bergin}, {Chandler}, {Masqu{\'e}}, {Maret}, \& {Heitsch}}]{tobin2011}
{Tobin}, J.~J., {Hartmann}, L., {Chiang}, H.-F., {et~al.} 2011,
  \bibinfo{title}{{Complex Structure in Class 0 Protostellar Envelopes. II.
  Kinematic Structure from Single-dish and Interferometric Molecular Line
  Mapping},} \apj, 740, 45, \dodoi{10.1088/0004-637X/740/1/45}

% type= article
\bibitem[{{\L}. {Tychoniec} {et~al.}(2024){Tychoniec}, {van Gelder}, {van
  Dishoeck}, {Francis}, {Rocha}, {Caratti o Garatti}, {Beuther}, {Gieser},
  {Justtanont}, {Linnartz}, \& et~al.}]{tychoniec2024}
{Tychoniec}, {\L}., {van Gelder}, M.~L., {van Dishoeck}, E.~F., {et~al.} 2024,
  \bibinfo{title}{{JWST Observations of Young protoStars (JOYS). Linked
  accretion and ejection in a Class I protobinary system},} \aap, 687, A36,
  \dodoi{10.1051/0004-6361/202348889}

% type= article
\bibitem[{R.~K. {Ulrich}(1976){Ulrich}}]{ulrich1976}
{Ulrich}, R.~K. 1976, \bibinfo{title}{{An infall model for the T Tauri
  phenomenon.},} \apj, 210, 377, \dodoi{10.1086/154840}

% type= article
\bibitem[{M.~L. {van Gelder} {et~al.}(2024){van Gelder}, {Francis}, {van
  Dishoeck}, {Tychoniec}, {Ray}, {Beuther}, {Caratti o Garatti}, {Chen},
  {Devaraj}, {Gieser}, \& et~al.}]{vanGelder2024}
{van Gelder}, M.~L., {Francis}, L., {van Dishoeck}, E.~F., {et~al.} 2024,
  \bibinfo{title}{{JWST Observations of Young protoStars (JOYS): Overview of
  gaseous molecular emission and absorption in low-mass protostars},} \aap,
  692, A197, \dodoi{10.1051/0004-6361/202451967}

% type= article
\bibitem[{M.~L.~R. {van't Hoff} {et~al.}(2023){van't Hoff}, {Tobin}, {Li},
  {Ohashi}, {J{\o}rgensen}, {Lin}, {Aikawa}, {Aso}, {de Gregorio-Monsalvo},
  {Gavino}, {Han}, {Koch}, {Kwon}, {Lee}, {Lee}, {Looney}, {Narayanan},
  {Plunkett}, {Sai}, {Santamar{\'\i}a-Miranda}, {Sharma}, {Sheehan},
  {Takakuwa}, {Thieme}, {Williams}, {Lai}, {Phuong}, \& {Yen}}]{van'thoff2023}
{van't Hoff}, M. L.~R., {Tobin}, J.~J., {Li}, Z.-Y., {et~al.} 2023,
  \bibinfo{title}{{Early Planet Formation in Embedded Disks (eDisk). III. A
  First High-resolution View of Submillimeter Continuum and Molecular Line
  Emission toward the Class 0 Protostar L1527 IRS},} \apj, 951, 10,
  \dodoi{10.3847/1538-4357/accf87}

% type= article
\bibitem[{P. Virtanen {et~al.}(2020)Virtanen, Gommers, Oliphant, Haberland,
  Reddy, Cournapeau, Burovski, Peterson, Weckesser, Bright, {van der Walt},
  Brett, Wilson, Millman, Mayorov, Nelson, Jones, Kern, Larson, Carey, Polat,
  Feng, Moore, {VanderPlas}, Laxalde, Perktold, Cimrman, Henriksen, Quintero,
  Harris, Archibald, Ribeiro, Pedregosa, {van Mulbregt}, \& {SciPy 1.0
  Contributors}}]{scipy2020}
Virtanen, P., Gommers, R., Oliphant, T.~E., {et~al.} 2020,
  \bibinfo{title}{{{SciPy} 1.0: Fundamental Algorithms for Scientific Computing
  in Python},} Nature Methods, 17, 261, \dodoi{10.1038/s41592-019-0686-2}

% type= article
\bibitem[{D.~M. {Watson} {et~al.}(2025){Watson}, {Narang}, {Pittman}, {Tyagi},
  {Gutermuth}, {Rubinstein}, {Evans}, {Hartmann}, {Megeath}, {Manoj}, \&
  et~al.}]{watson2025}
{Watson}, D.~M., {Narang}, M., {Pittman}, C.~V., {et~al.} 2025,
  \bibinfo{title}{{IPA. Accretion rate of a low-mass Class 0 protostar,
  measured via mid-infrared fluorescent OH emission},} arXiv e-prints,
  arXiv:2512.15999, \dodoi{10.48550/arXiv.2512.15999}

% type= article
\bibitem[{G.~S. {Wright} {et~al.}(2023){Wright}, {Rieke}, {Glasse}, {Ressler},
  {Garc{\'\i}a Mar{\'\i}n}, {Aguilar}, {Alberts}, {{\'A}lvarez-M{\'a}rquez},
  {Argyriou}, {Banks}, \& et~al.}]{wright2023}
{Wright}, G.~S., {Rieke}, G.~H., {Glasse}, A., {et~al.} 2023,
  \bibinfo{title}{{The Mid-infrared Instrument for JWST and Its In-flight
  Performance},} \pasp, 135, 048003, \dodoi{10.1088/1538-3873/acbe66}

\end{thebibliography}
\bibliographystyle{aasjournalv7}



\end{document}